\documentclass[draft, onecolumn, 12pt]{IEEEtran}

\usepackage{cite}

\ifCLASSINFOpdf
   \usepackage[final]{graphicx}
\else
\fi

\usepackage[cmex10]{amsmath}

\usepackage{amssymb}
\usepackage{pgfplots}
\pgfplotsset{compat=1.3}
\begin{document}
%
% paper title
% Titles are generally capitalized except for words such as a, an, and, as,
% at, but, by, for, in, nor, of, on, or, the, to and up, which are usually
% not capitalized unless they are the first or last word of the title.
% Linebreaks \\ can be used within to get better formatting as desired.
% Do not put math or special symbols in the title.
\title{Higher Order Statistics in\\Switched Diversity Systems}
%
%
% author names and IEEE memberships
% note positions of commas and nonbreaking spaces ( ~ ) LaTeX will not break
% a structure at a ~ so this keeps an author's name from being broken across
% two lines.
% use \thanks{} to gain access to the first footnote area
% a separate \thanks must be used for each paragraph as LaTeX2e's \thanks
% was not built to handle multiple paragraphs
%

\author{Adri\'{a}n Sauco-Gallardo,
        Unai Fern\'{a}ndez-Plazaola,
        Luis D\'{i}ez,
        and Eduardo Martos-Naya% <-this % stops a space
\thanks{This work is partially supported by the Spanish Government and FEDER under project TEC2011-25473 and the Junta de Andaluc\'{i}a under projects P11-TIC-7109 and P11-TIC-823.}% <-this % stops a space
\thanks{The authors are with Departamento de Ingenier\'{i}a de Comunicaciones, Universidad de M\'{a}laga
- Campus de Excelencia Internacional Andaluc\'{i}a Tech., M\'{a}laga 29071, Spain (e-mail: asaucog@ic.uma.es).}}% <-this % stops a space
%\thanks{Manuscript received April 19, 2005; revised September 17, 2014.}}

% note the % following the last \IEEEmembership and also \thanks - 
% these prevent an unwanted space from occurring between the last author name
% and the end of the author line. i.e., if you had this:
% 
% \author{....lastname \thanks{...} \thanks{...} }
%                     ^------------^------------^----Do not want these spaces!
%
% a space would be appended to the last name and could cause every name on that
% line to be shifted left slightly. This is one of those "LaTeX things". For
% instance, "\textbf{A} \textbf{B}" will typeset as "A B" not "AB". To get
% "AB" then you have to do: "\textbf{A}\textbf{B}"
% \thanks is no different in this regard, so shield the last } of each \thanks
% that ends a line with a % and do not let a space in before the next \thanks.
% Spaces after \IEEEmembership other than the last one are OK (and needed) as
% you are supposed to have spaces between the names. For what it is worth,
% this is a minor point as most people would not even notice if the said evil
% space somehow managed to creep in.

% make the title area
\maketitle

% As a general rule, do not put math, special symbols or citations
% in the abstract or keywords.
\begin{abstract}
We analyze the level crossing rate (LCR) and the average fade duration of the output signal-to-noise-ratio (SNR) in generalized switched diversity systems. By using a common approach, we study these higher order statistics for two different kinds of configurations: (1) Colocated diversity, i.e. receiver equipped with multiple antennas, and (2) Distributed diversity, i.e. relaying link with multiple single-antenna threshold-based decode-and-forward (DF) relays. In both cases, we consider the switched diversity combining strategies Selection Combining and Switch \& Stay Combining (SSC). Whenever using threshold-based techniques such as DF or SSC, the output SNR is a discontinuous random process and hence classic Rice approach to calculate the LCR is not applicable. Thus, we use an alternative formulation in terms of the one and two-dimensional  cumulative distribution functions of the output SNR. Our results are general, and hold for any arbitrary distribution of fading at the different diversity branches. Moreover, we develop a general asymptotic framework to calculate these higher order statistics in high mean SNR environments which only needs of the univariate probability density function.
\end{abstract}

% Note that keywords are not normally used for peerreview papers.

% For peer review papers, you can put extra information on the cover
% page as needed:
% \ifCLASSOPTIONpeerreview
% \begin{center} \bfseries EDICS Category: 3-BBND \end{center}
% \fi
%
% For peerreview papers, this IEEEtran command inserts a page break and
% creates the second title. It will be ignored for other modes.
\IEEEpeerreviewmaketitle

\section{Introduction}
% The very first letter is a 2 line initial drop letter followed
% by the rest of the first word in caps.
% 
% form to use if the first word consists of a single letter:
% \IEEEPARstart{A}{demo} file is ....
% 
% form to use if you need the single drop letter followed by
% normal text (unknown if ever used by IEEE):
% \IEEEPARstart{A}{}demo file is ....
% 
% Some journals put the first two words in caps:
% \IEEEPARstart{T}{his demo} file is ....
% 
% Here we have the typical use of a "T" for an initial drop letter
% and "HIS" in caps to complete the first word.
\IEEEPARstart{S}{witched} diversity techniques are trending upward since the introduction of distributed cooperative diversity \cite{coopdiversity1,coopdiversity2,coopdiversity3}. In this new setup, a source and a destination are willing to communicate with the help of $N$ single-antenna relays. Hence, these intermediate relay stations are regarded as a distributed agent, and different combining strategies are employed. These strategies are inspired on their single-link multi-antenna receiver counterparts\footnote{For the sake of clarity, we will refer to the single-link multi-antenna diversity counterpart as colocated diversity using the notation introduced in \cite{dsscperf5}.}, allowing for high-performance with low-complexity terminal equipment. Switched techniques arouse an special interest in this context due to their capability of achieving full-order spatial diversity \cite{switcheddiversity1,switcheddiversity2,switcheddiversity3,switcheddiversity4}, while all-participate strategies entailed loss of spectral efficiency as they require for orthogonal transmission.

Selection combining (SC) consists in switching each time to the best branch available, usually in terms of signal-to-noise-ratio (SNR), to perform the communication between source and destination \cite[Section 9.8]{alouini}. In the classical SC colocated multi-antenna receiver, selecting the branch with the best instant SNR implies that the $N$ available branches need to continuously be monitored. For this reason, Switch \& Stay Combining (SSC) was introduced as a way to avoid this requirement while allowing for an even simpler receiver. In SSC, a given branch is selected as long as the SNR on that branch remains above a given threshold \cite[Section 9.9]{alouini}. Nevertheless, as technology has developed, monitoring the SNR on every antenna is no more an issue and SSC had been left in low esteem until the appearance of the distributed diversity concept.

Since the introduction of distributed cooperative diversity, counterparts of these classic combining strategies can be found in recent literature. Distributed SC is known since its  introduction in \cite{csifb1} as \emph{Opportunistic Relaying} (OR). When considering OR, monitoring the SNR on every branch is no longer a matter of equipment's hardware complexity, but of amount feedback required to monitor the end-to-end SNR through every branch and to inform each relaying node whether it must retransmit or remain idle. As a result of this concern, SSC experiences a renewal of interest in distributed diversity schemes, since feedback is only required to request a switch to another relaying branch. To the best of our knowledge, the concept of distributed SSC was first introduced in \cite{dsscintro1,dsscintro2}. These distributed counterparts have fostered numerous works devoted to characterize the performance as much of OR, \cite{switcheddiversity3,switcheddiversity4,coopdiversity2,csifb1,aoraod}, as of distributed SSC systems \cite{dsscintro1,dsscintro2,dsscperf1,dsscperf3,dsscperf4,dsscperf5,dsscperf6}. Despite this evident interest, most analyses are focused on first order metrics such as the outage probability.

Second order statistics such as the level crossing rate (LCR) or the average fade duration (AFD) provide useful information related with the rate of change of a stationary random process, thus providing additional information about the dynamics of the process. Specifically, the LCR denotes how often the random process crosses a given threshold, whereas the AFD states the average amount of time that the random process remains below that threshold level. The knowledge of these second order statistics finds a variety of applications in the modeling and design of wireless communication systems, such as the amount of feedback required to successfully perform the communication or the latency of a link. The pioneering work by Rice \cite{rice} has allowed the computation of these metrics in a very general fashion, in terms of the joint distribution of the random process and its time derivative. One of the advantages of Rice's approach is its flexibility, as apparently mild conditions are imposed to the random process: stationarity, continuity, and smoothness of the correlation coefficient.

While these conditions hold for most common fading scenarios and hence the LCR and AFD have been already studied for different fading distributions in diversity combining scenarios \cite{ref_LCRMIMO_1,ref_LCRMIMO_2,ref_LCRMIMO_3,aoraod,letteraoraod}, the assumption of continuity in processes derived from threshold-based techniques, such as threshold-based decode-and-forward (DF) or SSC, is not realistic and hence Rice's framework finds an obstacle: the output SNR of dual-hop link with DF relaying and also the output SNR of a SSC scheme, either colocated or distributed, are inherently discontinuous. For this reason, analyses dealing with the LCR characterization of threshold-based techniques are scarce. To the best of our knowledge, the LCR characterization of the SSC scenario has only been studied in \cite{lcrssc1} under zero temporal correlation assumption using Rice's approach. Its results can be regarded as approximated: their accuracy was not corroborated by simulation and its plotted curves do not follow the shape of ours, which are corroborated by simulation. Nevertheless, the threshold-based DF obstacle has been recently circumvented by overlapping the results obtained from Rice's approach in different continuous crossing events \cite{aoraod}.

In this paper, we calculate in a general fashion the LCR and the AFD of processes derived from switching combining techniques using an alternative approach for calculating the higher order statistics of sampled processes \cite{paperedu}, which does not impose continuity as a condition. Furthermore, DF and switched diversity combining systems are implemented in a discrete time fashion, hence our processes are sampled, thus this other approach is the suitable one for our interest. We express the LCR and the AFD in terms of the univariate and bivariate CDF of the output SNR.

We investigate the SC, OR and SSC techniques in arbitrary fading conditions in this sampled fashion. Our results are valid both for the colocated and distributed configurations, and hold for any \textit{arbitrary} fading distribution. We regard the independent and identically distributed (IID) branches and independent non-identically distributed (InID) case, allowing for \textit{arbitrary} temporal correlation model of the per-branch fading for every scenario except for the colocated and distributed InID SSC case, where unfortunately we can only consider zero temporal correlation, i.e. independence between samples of the branch fading envelope. Furthermore, through our analysis, we find that the SSC technique in time-correlated fading environments benefits from having more than two diversity sources, against the usual recommendation which states that it does not \cite{sscsystemmodel}.

Finally, but not less interesting, we also derive simple approximated expressions in terms of the fading univariate probability density function (PDF) of the LCR and AFD in high mean SNR fading environments. That is, we achieve to accurately calculate this higher order statistics for SNR levels much lower than its mean value by only using the one-dimensional PDF of the process.

The remainder of this paper is organized as follows. Section \ref{systemmodel} introduces the scenarios of interest. In Section \ref{analisis} we present the general analysis for the LCR and AFD in the investigated scenarios. Then, in Section \ref{asintotico}, we discuss the asymptotic behavior of the LCR and AFD for high mean SNR. Section \ref{numericalresults} shows  numerical results which give validity to our theoretical expressions. Finally, the conclusions of our work are discussed in Section \ref{conclusiones}.

\section{System Model}\label{systemmodel}
\begin{figure}[!htb]
\centering
\includegraphics[width=16cm]{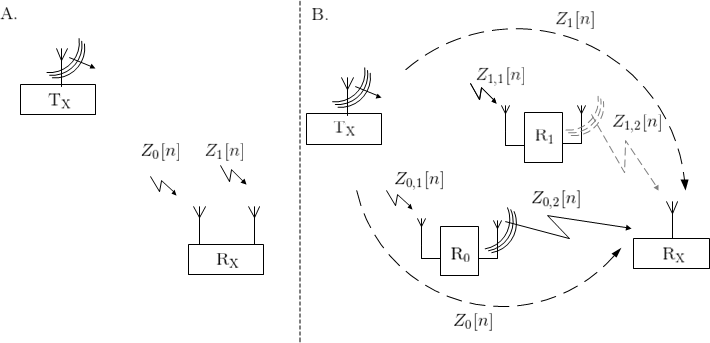}
\caption{A. Colocated diversity scenario. B. Distributed diversity scenario.\label{colocateddistributeddiversity}}
\end{figure}

We will consider two different spacial diversity configurations, regarded as colocated diversity and distributed diversity. Colocated diversity is the conventional multi-antenna receiver \cite{dsscperf5}, where the diversity sources are gathered together at the receiver; thus they are fully available all the time\footnote{We will discuss the diversity sources availability at the receiver side further on.}. Figure \ref{colocateddistributeddiversity}.A shows a colocated diversity scenario with two branches, $N=2$, where $Z_i[n]$ with $i\in [0\dots N-1]$ represents the SNR in the $n$th time interval on the $i$th branch, i.e. $i$th antenna. In the distributed case, for the sake of spectral efficiency, the diversity sources are not available all the time at the receiver side, hence the receiver must request them to the relay stations. Figure \ref{colocateddistributeddiversity}.B shows a distributed diversity scenario with two branches, $N=2$, where $Z_{i,1}[n]$, $Z_{i,2}[n]$ with $i\in [0\dots N-1]$ represent the SNR on the first and second hop of the $i$th branch, respectively, and $Z_i[n]$ with $i\in [0 \dots N-1]$ is the overall SNR through the $i$th branch, i.e. $i$th relayed channel. In both configurations SC or SSC is performed to get a system output with SNR $Z[n]$.

On each of these scenarios, we assume that additive white gaussian noise (AWGN) affects every branch, and that the stationary processes that characterize the different links are independent. We also consider as in \cite{aoraod} that switching from one branch to another is to be performed every time interval $T_S$,  short enough so that the instantaneous SNR on each antenna remains constant through $T_S$, nevertheless it will change from an interval to the next one. Note that we do not assume yet that consecutive samples from each branch are independent. In fact, for the SC/OR analysis we will not assume any kind of fading distribution nor correlation model, allowing their arbitrariness. For the SSC, analysis in the IID branches scenario we will allow for arbitrary fading distribution with any correlation model, however we will have to take the additional assumption of independent samples over time for characterizing the specific case of SSC with InID branches. This specific no correlation over time behavior can be found in time-division multiple-access (TDMA) systems, where the time interval between samples, $T_{TDMA}=L\times T_S$ is assumed to be significantly greater than the channel coherence time, where $L$ is the number of TDMA communications \cite{sscabc}. The SNR of each time interval is to be evaluated at its beginning so the chosen combining strategy can be performed based on this information and the availability of the diversity sources.

\subsection{Selection Combining}\label{scmodel}

SC combiner chooses each time the branch with the highest SNR. This technique benefits from fitting for both coherent and noncoherent modulation schemes, since the output of the SC combiner is equal to the signal on only one of the branches, hence the coherent sum of the individual branch signals is not required \cite[Section 9.8]{alouini}.

\subsubsection{Colocated Selection Combining}

As the diversity sources are provided by several antennas at the destination, these are all permanently available at the receiver side. Thus, at the beginning of each time interval the SNR on each branch is evaluated and the one with the greatest one is selected to demodulate the message. In terms of the output SNR, this scheme can be expressed as

\begin{equation}\label{SNRSC}
	Z[n]=\max (Z_i[n])
\end{equation}

\subsubsection{Opportunistic Relaying}

This term is usually found in literature for the SC distributed version. In this scheme, the destination has only one antenna and achieves spatial diversity through several relays. All relays listen to the source and attempt to decode the information, but only the one with the highest overall SNR $Z_i[n]$ retransmits to the destination. In \cite{csifb1}, it is described a procedure which achieves to make retransmit to the destination only to the relay with the highest overall SNR in each time interval without loss of spectral efficiency. Thus, given the availability of all the diversity sources all the time at the receiver, the output SNR can be expressed as in (\ref{SNRSC}) too.

\subsection{Switch \& Stay Combining}

SSC combiner chooses one of the diversity branches and stays with it as long as the SNR on it remains above a preset threshold $T$. When this occurs, the combiner switches to the another branch. As it does not use signals from different branches together at one time, this technique also benefits from fitting for both coherent and noncoherent modulation schemes as SC \cite[Section 9.9]{alouini}. The distribution of the output SNR will be given in terms of the distribution of the SNR on each input branch, but it will also depend on the instant the switching is to be performed, after switching condition is met, and on the order in which the different branches are selected. Because of this, we will consider that the system switching order is defined by a circular list of eligible branches, and that two different switching modes will be taken into account, each one related to the colocated and distributed scenarios respectively:

\subsubsection{Colocated Switch \& Stay Combining}
This is the classic manner of understanding SSC, as it is regarded in  \cite[Section 9.9.1]{alouini}. A instant switching (IS) policy, i.e. the switching is performed at the precise time interval when the switching event occurs, is appropriate for colocated diversity systems, such as a multiple antenna receiver, since the next diversity source to be switched to (i.e., the next receive antenna to select) will be available at that very interval. Attending to IS policy, the output SNR of a colocated SSC system ca be expressed as

\begin{equation*}
\begin{split}
&\begin{array}{lcr}
\mbox{\textbf{if}} & Z_s[n] < T &  \mbox{\textbf{then}}\\
&s=\langle s+1 \rangle&
\end{array}\\
&\mbox{\textbf{end if}}\\
&Z[n]=Z_s[n] 
\end{split}
\end{equation*}

where $s \in[0\dots N-1] $ represents the branch that the combiner is tracking; $Z[n]$ is the combiner output SNR in the $n$th interval; $T$ is the switching threshold; and $\langle \cdot \rangle$ represents the modulo operation, $\text{mod}(\cdot,N)$, which performs the circular step forward through the list of $N$  eligible branches.

\subsubsection{Distributed Switch \& Stay Combining}
We understand as a distributed SSC system as the one which achieves its diversity by being able to link through several relaying nodes as it is regarded in \cite{dsscintro1,dsscintro2}. A deferred switching (DS) policy, i.e. the switching is to be performed in the next time interval to the one on which the switching event actually occurs, is the suitable manner to perform SSC in a distributed system, where only one branch is available on each time interval, i.e. destination only links through a relaying node in each time interval, and the switching must be requested after the switching event occurrence. That is the destination will not detect the drop in the selected branch SNR until it receives it and must wait for the next time interval to request the link through another relaying node. Attending to DS policy, the output SNR of a colocated SSC system ca be expressed as

\begin{equation*}
\begin{split}
&Z[n]=Z_s[n]\\
&\begin{array}{lcr}
\mbox{\textbf{if}} & Z_s[n] < T &  \mbox{\textbf{then}}\\
&s=\langle s+1 \rangle&
\end{array}\\
&\mbox{\textbf{end if}}
\end{split}
\end{equation*}

\subsection{Dual-hop relaying in distributed scenarios}\label{canaltwohop}

In this subsection, we specialize our distributed configuration for a dual-hop threshold-based DF strategy. As in \cite{switcheddiversity4} or \cite{csifb1}, each time interval $T_S$ is divided in two slots: during the first slot, the relays are overhearing from the source; in the second slot, the source does not transmits and waits for the selected relay, based on the switched combining strategy decision, to retransmit to the destination. Specifically the investigated scenario operates as follows: for the first hop, when the received SNR at the relay is greater than a decoding threshold $T_{DF}$, noise free decodification is assumed and a perfect reconstruction of the transmitted signal is sent from the relay to the receiver. Thus, the distribution of the SNR at the receiver side only depends on the second hop link. When the received SNR at the relay is below $T_{DF}$, the relay does not decode and transmits nothing; thus the SNR at the receiver side will be identically $0$. As stated before, we assume the noise and fading caused by the first and second hop to be independent.

This mode of operation corresponds mathematically in terms of SNR to
\begin{equation}\label{snr_dfrelay}
Z_i[n] = \left\{ \begin{array}{rcl}
0 & \mbox{if} &  Z_{i,1}[n]<T_{DF}\text{,}\\
Z_{i,2}[n] & \mbox{if} & Z_{i,1}[n]\geq T_{DF}\text{.}

\end{array}\right.
\end{equation}

\section{Derivation of LCR and AFD}\label{analisis}
In order to analyze higher order statistics of the output SNR of switched diversity combining schemes, we will use the alternative approach used in \cite{paperedu}. Specifically, the LCR and the AFD can be obtained as:
\begin{align}
N_{Z}(u)&=\frac{F_{Z}(u)-F_{\mathbf{Z}}(u,u)}{T_S}\text{,}\label{lcr}\\
A_{Z}(u)&=\frac{F_{Z}(u)}{N_{Z}(u)}\text{,}\label{afd}
\end{align}
where $N_{Z}(u)$ is the LCR of the stationary process $Z[n]$ through the level $u$; $A_{Z}(u)$ is the AFD of $Z[n]$; $F_{Z}(\cdot)$ is the CDF of $Z[n]$; $F_{\mathbf{Z}}(\cdot,\cdot)$ is the bivariate CDF of the vector of two consecutive samples of $Z[n]$, $\mathbf{Z}=\bigl[ Z[n],Z[n+1]\bigr]$; and $T_S$ is the time interval between samples.

As \eqref{lcr} and \eqref{afd} show, only the univariate and bivariate CDF of the output are needed to calculate its LCR and AFD. From now on, we will focus on obtaining the analytical expressions for these CDFs.

\subsection{Selection Combining}

As stated in section \ref{scmodel}, the output SNR can be express in terms of the per-branch SNR the same way for the colocated SC scenario or the distributed OR scenario. Thus, the expressions for the univariate and bivariate CDF are the same. Specifically, we can write the SC/OR output SNR univariate CDF in terms of the per-branch SNR CDFs as

\begin{equation}\label{scunivariate}
\begin{split}
F_Z(u)&=\text{Pr}\{ \max(Z_i[n]) < u \}=\text{Pr}\{ Z_0[n] < u, Z_1[n] < u\dots Z_N[n]<u \}=\prod^{N-1}_{i=0}F_{Z_i}(u)\text{,}
\end{split}
\end{equation}

where we have made use of the independence between fading links assumption and $F_{Z_i}(\cdot)$ is the univariate CDF of the SNR on the $i$th branch. In the same fashion, we express the output SNR bivariate CDF as

\begin{equation}\label{scbivariate}
\begin{split}
F_\mathbf{Z}(u_1,u_2)&=\text{Pr}\{ \max(Z_i[n]) < u_1, \max(Z_i[n+1]) < u_2 \} \\
&\begin{split}
=\text{Pr}\{ &Z_0[n] < u_1\dots Z_N[n]<u_1,Z_0[n+1] < u_2\dots Z_N[n+1]< u_2 \}
\end{split}\\
&=\prod^{N-1}_{i=0}F_{\mathbf{Z}_i}(u_1,u_2)\text{,}
\end{split}
\end{equation}

where $F_{\mathbf{Z}_i}(\cdot,\cdot)$ is the bivariate CDF of the SNR of two consecutive samples $Z_i[n], Z_i[n+1]$ on the $i$th branch.

Given the expressions in (\ref{scunivariate}) and (\ref{scbivariate}), the LCR and AFD of the output SNR of any SC/OR system can be easily computed when expressions of the per-branch SNR univariate and bivariate CDF are available.

\subsection{Switch \& Stay Combining}

We will first calculate expressions for the CDFs of interest, in terms of the distributions of the SNRs on each branch. For the colocated and distributed SSC cases, we will analyze two different scenarios: 

\begin{itemize}
	\item \emph{IID-AC}: IID fading with arbitrary distribution and arbitrary temporal correlation model.
	\item \emph{InID-TI}: InID fading with arbitrary distributions and temporal independence between its samples.
\end{itemize}

In general, the univariate and bivariate CDF of the SNR level at the output of the combiner can be expressed as
\begin{equation}
F_Z(u)=\sum^{N-1}_{i=0}P_i F_{Z|s=i}(u),
\end{equation}
\begin{equation}
F_{\mathbf{Z}}(u_1,u_2)=\sum^{N-1}_{i=0}P_i F_{\mathbf{Z}|s=i}(u_1,u_2),
\end{equation}
where $P_i$ is the probability of the combiner being tracking the $i$th branch at any given time and $F_{Z|s=i}(u)$, $F_{\mathbf{Z}|s=i}(u_1,u_2)$ are respectively the univariate and bivariate CDF of the output SNR level, conditioned on the combiner being tracking the $i$th branch at the beginning of the observation interval.

For the IID-AC case, it is trivial that $P_i=N^{-1}$, $F_{Z|s=i}(u)=F_{Z|s=j}(u)$ and $ F_{\mathbf{Z}|s=i}(u_1,u_2)= F_{\mathbf{Z}|s=j}(u_1,u_2)$ $\forall i,j \in [0\dots N-1]$.

For the InID-TI case, an expression for $P_i$ can be found in \cite[Equation 2]{sscsystemmodel} as
\begin{equation}
P_i=\Biggl( F_{Z_i}(T) \sum^{N-1}_{k=0}\frac{1}{ F_{Z_k}(T)} \Biggr)^{-1}\text{.}
\end{equation}

Unfortunately, to the best of our knowledge, there is not an expression for $P_i$ in a SSC InID scenario which allows for correlation over time on each branch.

Now we proceed to find the expressions of  $ F_{Z|s=i}(u)$, $F_{\mathbf{Z}|s=i}(u_1,u_2)$ for each of the investigated switching modes.

\subsubsection{Colocated SSC}

Assuming IS is performed, the combiner output SNR univariate CDF, conditioned on the combiner being tracking the $i$th branch at the beginning of the observation interval, reduces to an expression similar to \cite[Eq. 9.270]{alouini},
\begin{equation}\label{is-ssc_univariate}
F_{Z|s=i}(u) = \left\{ \begin{array}{lcl}
F_{Z_i}(T)\times F_{Z_{\langle i+1 \rangle}}(u) & \mbox{for} & u<T\text{,} \\
\\
\begin{split}
&F_{Z_i}(T)\times F_{Z_{\langle i+1 \rangle}}(u)+F_{Z_i}(u) - F_{Z_i}(T)
\end{split} & \mbox{for} & u\geq T\text{.}
\end{array}\right.
\end{equation}

On the other hand, the combiner output SNR bivariate CDF expression is divided in four intervals, $u_1\text{,}u_2<T$; $u_2<T\leq u_1$; $u_1<T\leq u_2$; and $u_1\text{,}u_2\geq T$. By defining
\begin{equation}\label{psibivarCDF}
\Psi^{(i)}_N (u_1,u_2) = \left\{ \begin{array}{lcl}
F_{\mathbf{Z}_i}(T,u_2)\times F_{\mathbf{Z}_{\langle i+1 \rangle}}(u_1,T) & \mbox{for} & N=2\text{,} \\
\\
\begin{split}&F_{Z_i}(T)\times F_{\mathbf{Z}_{\langle i+1 \rangle}}(u_1,T)\times F_{Z_{\langle i+2 \rangle}}(u_2)\end{split} & \mbox{for} & N\geq 3\text{,}
\end{array}\right.
\end{equation}
the output SNR bivariate CDF corresponds to

\begin{equation}\label{is-ssc-bivariate}
F_{\mathbf{Z}|s=i}(u_1,u_2) = \left\{ \begin{array}{lcl}
\Psi^{(i)}_N(u_1,u_2) & \mbox{for} & u_1,u_2<T\text{,} \\
\\
\begin{split}
&\Psi^{(i)}_N(u_1,u_2)\\
&+ F_{Z_{\langle i+1 \rangle}}(u_2)\times \Bigl(F_{\mathbf{Z}_i}(u_1,T)-F_{\mathbf{Z}_i}(T,T)\Bigr)
\end{split} & \mbox{for} & u_2 < T \leq u_1\text{,}\\
\\
\begin{split}
&\Psi^{(i)}_N(u_1,u_2) \\
&+ F_{Z_i}(T)\times \Bigl(F_{\mathbf{Z}_{\langle i+1 \rangle}}(u_1,u_2)-F_{\mathbf{Z}_{\langle i+1 \rangle}}(u_1,T)\Bigr)
\end{split} & \mbox{for} & u_1 < T \leq u_2\text{,}\\
\\
\begin{split}
&\Psi^{(i)}_N(u_1,u_2) + F_{\mathbf{Z}_i}(u_1,u_2) - F_{\mathbf{Z}_i}(u_1,T)\\
&- F_{\mathbf{Z}_i}(T,u_2)+ F_{\mathbf{Z}_i}(T,T)\\
&+ F_{Z_{\langle i+1 \rangle}}(u_2)\times \Bigl(F_{\mathbf{Z}_i}(u_1,T)-F_{\mathbf{Z}_i}(T,T)\Bigr)\\
&+ F_{Z_i}(T)\times \Bigl(F_{\mathbf{Z}_{\langle i+1 \rangle}}(u_1,u_2)-F_{\mathbf{Z}_{\langle i+1 \rangle}}(u_1,T)\Bigr)
\end{split} & \mbox{for} & u_1,u_2 \geq T\text{.}
\end{array}\right.
\end{equation}

%\begin{figure*}[!t]
%
%
%
%\hrulefill
%\end{figure*}

We observe in (\ref{psibivarCDF}), (\ref{is-ssc-bivariate}) that the bivariate CDF is different for $N=2$ and $N\geq 3$; hence, the number of branches must be regarded when calculating statistics whose observation interval allows for the occurrence of two switchings. The expression must differ between a two-branch combiner, $N=2$, and a combiner with three or more branches, $N\geq 3$, since in the two-branch combiner case, after the second switch, the combiner returns to the original branch.

\subsubsection{Distributed SSC}

Assuming DS is performed, the combiner output SNR univariate CDF, conditioned on the combiner being tracking the $i$th branch at the beginning of the observation interval, can be expressed as
\begin{equation}\label{ssc_distributed_univariate}
F_{Z|s=i}(u)=F_{Z_i}(u)\text{.}
\end{equation}

The output CDF turns out to be the same CDF of the selected branch SNR, as the combiner outputs this branch regardless of its SNR level. The SNR level of the selected branch only determines the selected branch on the next interval.

In this case, the bivariate CDF only has two regions and does not depend on the number of branches, since only one switch can occur during the observation interval of this second order statistic:

\begin{equation}\label{ssc_distributed_bivariate}
\begin{split}
F_{\mathbf{Z}|s=i}&(u_1,u_2) = \left\{ \begin{array}{lcl}
F_{Z_i}(u_1)\times F_{Z_{\langle i+1 \rangle}}(u_2) & \mbox{for} & u_1<T\text{,} \\
\\
\begin{split}
&F_{\mathbf{Z}_i}(u_1,u_2)-F_{\mathbf{Z}_i}(T,u_2)+F_{Z_i}(T)\times F_{Z_{\langle i+1 \rangle}}(u_2)
\end{split} & \mbox{for} & u_1\geq T\text{.}
\end{array}\right.
\end{split}
\end{equation}

\subsection{Distributions on each branch}
We now settle the expressions of the univariate and bivariate CDF of the SNR on each branch, $F_{Z_i}(\cdot)$ and $F_{\mathbf{Z}_i}(\cdot,\cdot)$.

In the colocated diversity scenarios, this reduces to CDFs of each branch SNR. In the distributed scenarios, we assume a DF relay network operating as indicated  in section \ref{canaltwohop}. 

Considering that the SNR distribution on each hop is known, the overall SNR CDF on each branch can be expressed as
\begin{equation}\label{distributed_univariate}
\begin{split}
F_{Z_i}(u)=&F_{Z_{i,1}}(T_{DF}) + F_{Z_{i,2}}(u)\times\bigl( 1 - F_{Z_{i,1}}(T_{DF}) \bigr),
\end{split}
\end{equation}
where $F_{Z_{i,1}}(\cdot)$ and $F_{Z_{i,2}}(\cdot)$ are the CDFs of the SNR on the first and second hop of the $i$th branch, respectively.

The bivariate CDF of each branch in the distributed scenario can be expressed as
\begin{equation}\label{distributed_bivariate}
\begin{split}
F_{\mathbf{Z}_i}(u_1,u_2)=&F_{\mathbf{Z}_{i,1}}(T_{DF},T_{DF}) + \bigl( F_{Z_{i,2}}(u_1) + F_{Z_{i,2}}(u_2) \bigr)\times\bigl(F_{Z_{i,1}}(T_{DF}) - F_{\mathbf{Z}_{i,1}}(T_{DF},T_{DF})\bigr)\\
&+F_{\mathbf{Z}_{i,2}}(u_1,u_2)\times\bigl( 1 +  F_{\mathbf{Z}_{i,1}}(T_{DF},T_{DF})- 2\times F_{Z_{i,1}}(T_{DF})\bigr),
\end{split}
\end{equation}
where $F_{\mathbf{Z}_{i,1}}(\cdot,\cdot)$ and $F_{\mathbf{Z}_{i,2}}(\cdot,\cdot)$ are the bivariate CDF of the SNR on the first and second hop of the $i$th branch, respectively.

We must recall that when assuming independence between samples of a branch, every $F_{\mathbf{Z}_i}(u_1,u_2)=F_{Z_i}(u_1)\times F_{Z_i}(u_2)$

\section{High Mean SNR Asymptotic Behavior}\label{asintotico}

While the approach discussed in section \ref{analisis} is general and exact, it does not offer a solution when the bivariate CDF of the fadings involved in the scenario is not available. For this reason, and to provide some insight about the impact of the scenario parameters on studied higher order statistics, we proceed to develop simpler expressions which, only requiring the knowledge of the univariate PDF, describe considerably accurately the behavior of these higher order statistics in scenarios where the mean SNR is much greater than the level $u$.

\subsection{LCR in High Mean SNR Scenarios}

Let us note that the univariate and bivariate CDF of any fading SNR in addition to $u$ is also composed of a parameter which we had omitted before for the sake of compactness: the mean SNR, $\Omega$. Moreover, the correlation coefficient, $\rho$, has also been omitted from the bivariate CDF. Thus, $F_{Z}(u)\equiv F_{Z}(u|\Omega)$ and $F_{\mathbf{Z}}(u_1,u_2)\equiv F_{\mathbf{Z}}(u_1,u_2|\Omega,\rho)$. We define the normalized SNR, $\bar{Z}[n]=Z[n]/\Omega$, such as the expectation of $\bar{Z}$ is $1$, and its CDFs $F_{\bar{Z}}(u)$, $F_{\bar{\mathbf{Z}}}(u_1,u_2|\rho)$. By defining the normalized $u$, $\bar{u}=u/\Omega$ we can rewrite the LCR expression in (\ref{lcr}) as

\begin{equation}\label{lcr_znormalizada}
N_Z(\bar{u})=\frac{F_{\bar{Z}}(\bar{u})-F_{\bar{\mathbf{Z}}}(\bar{u},\bar{u}|\rho)}{T_S}.
\end{equation}

Hence, if equation (\ref{lcr_znormalizada}) is infinitely differentiable in $\bar{u}$, the LCR for high mean SNR environments can be expanded as a Taylor series at $\bar{u}=0$,

\begin{equation}
\begin{split}
N_Z&(\bar{u})=\frac{1}{T_S}\sum^\infty_{n=0}\frac{\bar{u}^n}{n!}\frac{d^n}{d\bar{u}^n}\bigl( F_{\bar{Z}}(\bar{u})- F_{\bar{\mathbf{Z}}}(\bar{u},\bar{u}|\rho) \bigr)\bigr|_{\bar{u}=0}.
\end{split}
\end{equation}

Typically, the univariate and bivariate CDF of a positive random variable is null at $0$. Thus, the first non-null term of the LCR Taylor series expansion at $\bar{u}=0$ is enough to characterize the behavior of the LCR when asymptotically $u/\Omega\rightarrow 0$. Nevertheless, we have introduced distributed scenarios with DF relaying where the end-to-end SNR can be identically null with a finite probability, making theirs univariate and bivariate CDF non-null at $0$. In this case, the asymptotic value of the LCR for low values of $\bar{u}$ is immediately the difference between the univariate and bivariate CDF divided by $T_S$. In the rest of cases, it is clear that the LCR tends to be null for low values of $\bar{u}$, but we attempt to know how it approaches to $0$. Thus, we study now the behavior of the $n$th derivatives of the univariate and bivariate CDF.

Using equation \cite[(0.410)]{tableintegrals}, the $n$th derivative of any univariate CDF can be expressed as

\begin{equation}
\frac{d^n}{d\bar{u}^n} F_{\bar{Z}}(\bar{u})=\frac{d^{(n-1)}}{d\bar{u}^{(n-1)}}f_{\bar{Z}}(\bar{u}),
\end{equation}

where $f_{\bar{Z}}(\bar{u})$ is the PDF of the normalized fading SNR, whereas the $n$th derivative of the bivariate CDF can be expressed as

\begin{equation}\label{nderivadabivariate}
\begin{split}
\frac{d^n}{d\bar{u}^n} F_{\bar{\mathbf{Z}}}(\bar{u},\bar{u}|\rho)=2\biggl( \int^{\bar{u}}_0 f_{\bar{Z}_2|\bar{Z}_1}(\bar{z}_2|\bar{Z}_1=\bar{u})d\bar{z}_2\times\frac{d^{(n-1)}}{d\bar{u}^{(n-1)}}f_{\bar{Z}}(\bar{u})+\sum^{n-2}_{k=0} \Phi_{n,k}(\bar{u},\rho)\frac{d^{k}}{d\bar{u}^{k}}f_{\bar{Z}}(\bar{u}) \biggr),
\end{split}
\end{equation}

where $f_{\bar{Z}_2|\bar{Z}_1}(\bar{z}_2|\bar{Z}_1=\bar{z}_1)$ is the PDF of a sample of the normalized SNR conditioned to the value of the previous sample, and $\Phi_{n,k}(\bar{u},\rho)$ gathers together functions which multiply lower order derivatives of the PDF to compose the expression of the $n$th derivative of the bivariate CDF.

Note now that, if we assume the continuity of conditioned PDF around $0$, the integral expression multiplying the highest order derivative of the PDF involved in (\ref{nderivadabivariate}) for $\bar{u}=0$, becomes $0$ due to the limits of integration. Hence, in the inspection process in search of the first non-null term, we will evaluate at $\bar{u}=0$ the successive derivatives of the PDF until we locate the first non-null. At this point, we will have the value of the univariate CDF term, whereas we will have ensure that the bivariate CDF term is still null, since the first non-zero derivative term is multiplied by zero and the lower order derivative terms have been proved to be zero.

If we call $m$ to order of the first non-null derivative of the LCR at $\bar{u}=0$, we can express the LCR in high mean SNR environments as

\begin{equation}
N_Z(\bar{u})|_{\bar{u}\rightarrow 0} \sim \frac{1}{m!T_S}\frac{d^{(m-1)}f_{\bar{Z}}(\bar{u})}{d\bar{u}^{(m-1)}}\biggr|_{\bar{u}=0}( \bar{u})^m,
\end{equation}

which we normalize to the sampling period, $\bar{N}_Z(\bar{u})=N_Z(\bar{u})T_S$, and express in logarithmic scale to find that

\begin{equation}
\log \bigl( \bar{N}_Z(\bar{u}) \bigr) |_{\bar{u}\rightarrow 0} \sim m\log(\bar{u}) + \log\biggl( \frac{1}{m!}\frac{d^{(m-1)}f_{\bar{Z}}(\bar{u})}{d\bar{u}^{(m-1)}} \biggr),
\end{equation}

the LCR in a high mean SNR scenario tends to a slope-intercept form where $m$, order of the first non-null derivative, determines the slope of the straight line asymptote; whereas the value of this $m$th derivative shifts left the asymptote. Thus we have obtained an simple expression for the asymptotic behavior of the LCR of a sampled process where only the knowledge of the PDF is required.

\subsection{AFD in High Mean SNR Scenarios}

After the asymptotic LCR analysis, recalling (\ref{afd}), assuming once again that the univariate and bivariate CDF of the process is null at $0$, we can express the asymptotic behavior of the AFD as

\begin{equation}
A_Z(\bar{u})|_{\bar{u}\rightarrow 0}\sim T_S,
\end{equation}

since the LCR asymptotic behavior in this case is the same to the univariate CDF asymptotic behavior. From this expression we observe that, for high mean SNR environments, the AFD tends to last the duration of a single sample of our discrete process, i.e. such fadings are so improbable that, in case of occurrence, its mean duration is the minimum possible.

Whereas, when the univariate or bivariate CDF are non-null at $0$, this $A_Z(u)|_{u/\Omega\rightarrow 0}$ is higher with direct proportion to how much probability is condensed at $0$. For that case, we can write

\begin{equation}
A_Z(\bar{u})|_{\bar{u}\rightarrow 0}\sim T_S\frac{F_Z(0)}{F_Z(0)-F_{\mathbf{Z}}(0,0)}.
\end{equation}

\section{Numerical Results}\label{numericalresults}

In this section, we use the derived expressions for the higher order statistics to evaluate their behavior in different scenarios: SC, OR, colocated SSC and distributed SSC in Rayleigh, Nakagami-m and Hoyt fading environments. For the sake of simplicity, we plot the exact curves for Rayleigh fading scenarios, whose bivariate CDF follows a simple expression in terms of the Marcum $Q$-function \cite[Eq.~6.5]{alouini}, verify them with simulation points, and compare them to the asymptotic straight line expression. While for the Nakagami-m and Hoyt scenarios we plot the low $u/\Omega$ asymptotic straight lines and compare them with simulated points. Note that exact calculation of other fading distributions can be easily considered by plugging readily available expressions for the bivariate CDFs. We assume that the underlying Gaussians related to each probability distribution model of each fading link experience the temporal correlation model proposed by Clarke \cite[Chapter\ 1]{jakes}. Thus, the SNR correlation of each link is defined by the coefficient
\begin{equation}
\rho=\frac{\text{cov}(Z[n],Z[n+1])}{\text{var}(Z)}=\bigl| J_0\bigl( 2\pi f_DT_S\bigr) \bigr|^2,
\end{equation}
where $\text{cov}(\cdot,\cdot)$ is the covariance, $\text{var}(\cdot)$ the variance, $J_0(\cdot)$ is the Bessel function of the first kind and order $0$ and $f_D$ is the Doppler frequency of the channel. For the following results we use the values $f_DT_S=0.05$, $0.2$ and $0.38$, which lead to $\rho\approx 0.95$, $0.41$ and $0$.

We use the levels $u$, $T$ and $T_{DF}$ in a normalized manner, $\bar{u}=u/\Omega$, $\bar{T}=T/\Omega$, $\bar{T}_{DF}=T_{DF}/\Omega$, where $\Omega$ is the mean SNR on each of the considered IID fading links.

\begin{figure}[htb!]
\centering
\pgfplotsset{every axis/.append style={
xmin={-20},
xmax={10},
ymax={1},
ymin={0.00001},
%extra description/.code={
%\node at (0.58,0.47) {\small{$\Delta T_{DF}$}};
%\node at (0.66,0.38) {\small{$\Delta T$}};
%}
}}
\begin{tikzpicture}[scale=1]
\begin{semilogyaxis}[
width=16cm,
height=16cm,
xlabel=$\bar{u}(\text{dB})$,
ylabel=$N_Z(\bar{u})\times T_S$,
grid=both,
legend entries={$\rho = 0.95$,$\rho = 0.41$,$\rho = 0$},
legend style={legend pos=south east}
]

\addplot[color=magenta, mark=*] table[x=x,y=y] {Legend.dat};
\addplot[color=cyan, mark=square*] table[x=x,y=y] {Legend.dat};
\addplot[color=orange, mark=triangle*] table[x=x,y=y] {Legend.dat};

\addplot[color=magenta] table[x=x,y=y] {a1.dat};
\addplot[color=cyan] table[x=x,y=y] {a2.dat};
\addplot[color=orange] table[x=x,y=y] {a3.dat};

\addplot[color=magenta] table[x=x,y=y] {a4.dat};
\addplot[color=cyan] table[x=x,y=y] {a5.dat};
\addplot[color=orange] table[x=x,y=y] {a6.dat};

\addplot[color=magenta, mark=*, only marks] table[x=x,y=y] {a7.dat};
\addplot[color=cyan, mark=square*, only marks] table[x=x,y=y] {a8.dat};
\addplot[color=orange, mark=triangle*, only marks] table[x=x,y=y] {a9.dat};

\addplot[color=magenta, mark=*, only marks] table[x=x,y=y] {a10.dat};
\addplot[color=cyan, mark=square*, only marks] table[x=x,y=y] {a11.dat};
\addplot[color=orange, mark=triangle*, only marks] table[x=x,y=y] {a12.dat};

\addplot[color=black, dash pattern=on 6pt off 2pt on 1pt off 2pt] table[x=x,y=y] {a13.dat};
\addplot[color=black, dash pattern=on 6pt off 2pt on 1pt off 2pt] table[x=x,y=y] {a14.dat};

\draw [stealth-] (axis cs:-12.8,0.0042) -- (axis cs:-13.5,0.011) node[above]{$N=2$};
\draw[] (axis cs:-12.5,0.0025) circle (0.6cm);

\draw [stealth-] (axis cs:-8.2,0.0008) -- (axis cs:-8.9,0.0018) node[above]{ $N=4$};
\draw[] (axis cs:-7.5,0.0005) circle (0.6cm);

\end{semilogyaxis}
\end{tikzpicture}
\caption{Normalized LCR of the output SNR in colocated SC combiners with $2$ and $4$ antennas in IID Rayleigh scenario.}\label{fig_csc_rayleigh}
\end{figure}
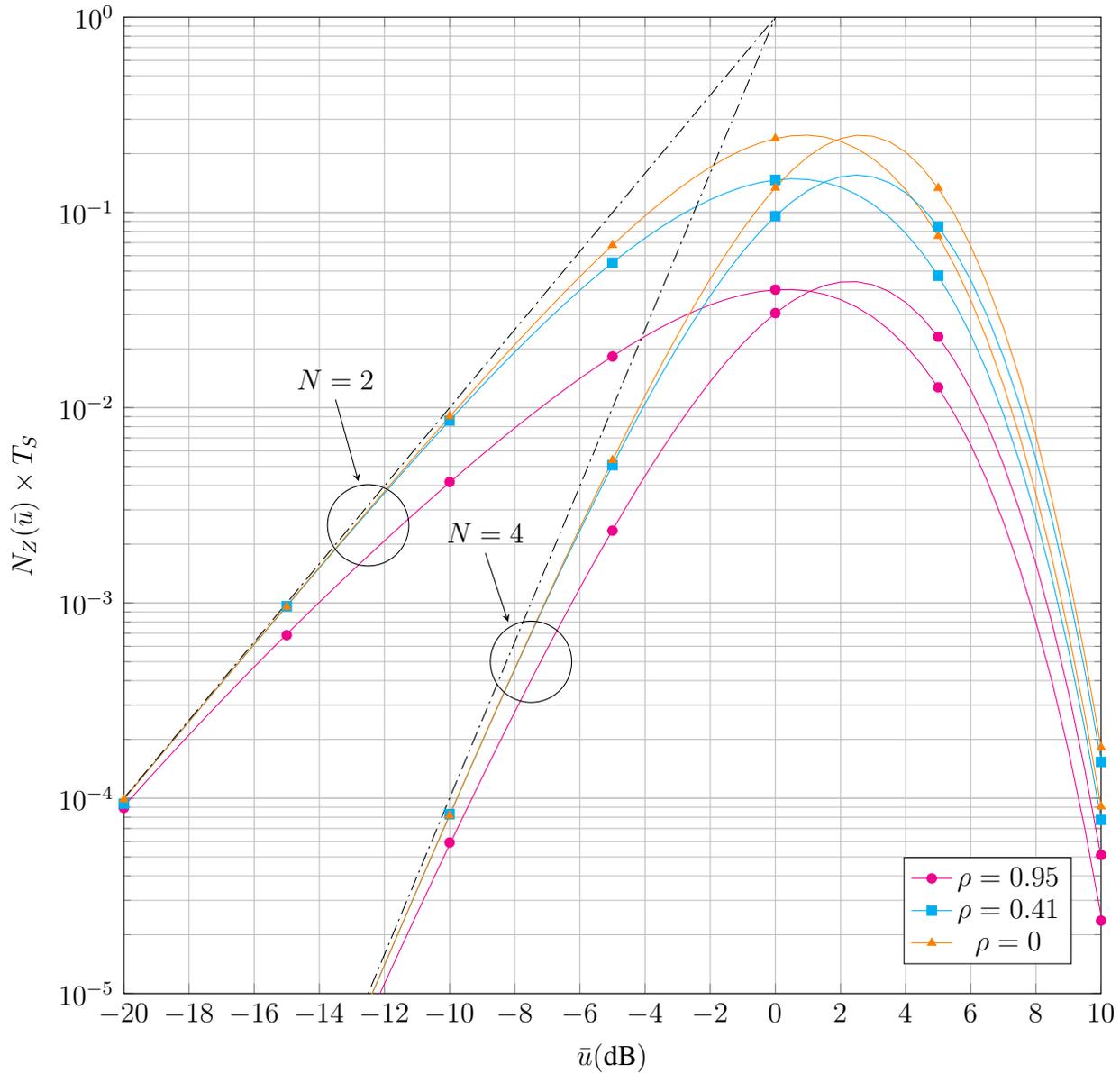

\begin{figure}[htb!]
\centering
\pgfplotsset{every axis/.append style={
xmin={-25},
xmax={8},
ymax={0.5},
ymin={0.0000005},
%extra description/.code={
%\node at (0.58,0.47) {\small{$\Delta T_{DF}$}};
%\node at (0.66,0.38) {\small{$\Delta T$}};
%}
}}
\begin{tikzpicture}[scale=1]
\begin{semilogyaxis}[
width=16cm,
height=16cm,
xlabel=$\bar{u}(\text{dB})$,
ylabel=$N_Z(\bar{u})\times T_S$,
grid=both,
legend entries={$\rho = 0.95$,$\rho = 0.41$,$\rho = 0$},
legend style={ legend pos=south east}
]

\addplot[color=magenta, densely dashed, mark=*] table[x=x,y=y] {Legend.dat};
\addplot[color=cyan, densely dotted, mark=square*] table[x=x,y=y] {Legend.dat};
\addplot[color=orange, dash pattern=on 3pt off 1pt on 1pt off 1pt, mark=triangle*] table[x=x,y=y] {Legend.dat};

%%%%%%%%%%%%%%%%%%%%%%%%%%%%%%%%%%%%%%%%%%%%%%%%%%%%%%%%%%%%%%%%%%%%%%%%%%%%%%%%%%%%%%%%%%%%%%%%%%%%%%%%%

\addplot[color=magenta, densely dashed, mark=*] table[x=x,y=y] {b1.dat};
\addplot[color=cyan, densely dotted, mark=*] table[x=x,y=y] {b2.dat};
\addplot[color=orange, dash pattern=on 3pt off 1pt on 1pt off 1pt, mark=triangle*] table[x=x,y=y] {b3.dat};

\draw [stealth-] (axis cs:-22,0.0012) -- (axis cs:-20.5,0.005) node[above]{$\begin{tabular}{c} N=2\text{,}\\ q=0.1 \end{tabular}$};
\draw[] (axis cs:-22.5,0.0007) circle (0.4cm);

%%%%%%%%%%%%%%%%%%%%%%%%%%%%%%%%%%%%%%%%%%%%%%%%%%%%%%%%%%%%%%%%%%%%%%%%%%%%%%%%%%%%%%%%%%%%%%%%%%%%%%%%%

\addplot[color=black, dash pattern=on 6pt off 2pt on 1pt off 2pt] table[x=x,y=y] {b4.dat};

\addplot[color=black, dash pattern=on 6pt off 2pt on 1pt off 2pt] table[x=x,y=y] {b5.dat};

%%%%%%%%%%%%%%%%%%%%%%%%%%%%%%%%%%%%%%%%%%%%%%%%%%%%%%%%%%%%%%%%%%%%%%%%%%%%%%%%%%%%%%%%%%%%%%%%%%%%%%%%%%%

\addplot[color=magenta, densely dashed, mark=*] table[x=x,y=y] {b6.dat};
\addplot[color=cyan, densely dotted, mark=square*] table[x=x,y=y] {b7.dat};
\addplot[color=orange, dash pattern=on 3pt off 1pt on 1pt off 1pt, mark=triangle*] table[x=x,y=y] {b8.dat};

\draw [stealth-] (axis cs:-22,0.000055) -- (axis cs:-20.5,0.00013) node[above]{$\begin{tabular}{c} N=2\text{,}\\ q=0.9 \end{tabular}$};
\draw[] (axis cs:-22.5,0.000033) circle (0.4cm);

%%%%%%%%%%%%%%%%%%%%%%%%%%%%%%%%%%%%%%%%%%%%%%%%%%%%%%%%%%%%%%%%%%%%%%%%%%%%%%%%%%%%%%%%%%%%%%%%%%%%%%%%%%

\addplot[color=magenta, densely dashed, mark=*] table[x=x,y=y] {b9.dat};
\addplot[color=cyan, densely dotted, mark=square*] table[x=x,y=y] {b10.dat};
\addplot[color=orange, dash pattern=on 3pt off 1pt on 1pt off 1pt, mark=triangle*] table[x=x,y=y] {b11.dat};

\draw [stealth-] (axis cs:-21.5,0.0000007) -- (axis cs:-18,0.0000011) node[above]{$\begin{tabular}{c} N=4\text{,}\\ q=0.1 \end{tabular}$};
\draw[] (axis cs:-22.5,0.0000008) circle (0.4cm);

%%%%%%%%%%%%%%%%%%%%%%%%%%%%%%%%%%%%%%%%%%%%%%%%%%%%%%%%%%%%%%%%%%%%%%%%%%%%%%%%%%%%%%%%%%%%%%%%%%%%%%%%%

\addplot[color=black, dash pattern=on 6pt off 2pt on 1pt off 2pt] table[x=x,y=y] {b12.dat};

\addplot[color=black, dash pattern=on 6pt off 2pt on 1pt off 2pt] table[x=x,y=y] {b13.dat};

%%%%%%%%%%%%%%%%%%%%%%%%%%%%%%%%%%%%%%%%%%%%%%%%%%%%%%%%%%%%%%%%%%%%%%%%%%%%%%%%%%%%%%%%%%%%%%%%%%%%%%%%%%%

\addplot[color=magenta, densely dashed, mark=*] table[x=x,y=y] {b14.dat};
\addplot[color=cyan, densely dotted, mark=square*] table[x=x,y=y] {b15.dat};
\addplot[color=orange, dash pattern=on 3pt off 1pt on 1pt off 1pt, mark=triangle*] table[x=x,y=y] {b16.dat};

\draw [stealth-] (axis cs:-14,0.0000008) -- (axis cs:-9.8,0.000002) node[above]{$\begin{tabular}{c} N=4\text{,}\\ q=0.9 \end{tabular}$};
\draw[] (axis cs:-15,0.0000008) circle (0.4cm);

\end{semilogyaxis}
\end{tikzpicture}
\caption{Comparison of simulation results and asymptotic behaviour of normalized LCR of the output SNR in a colocated SC combiners with $2$ and $4$ antennas in $q=0.1$ and $q=0.9$ IID Hoyt scenario.}\label{fig_csc_hoyt}
\end{figure}
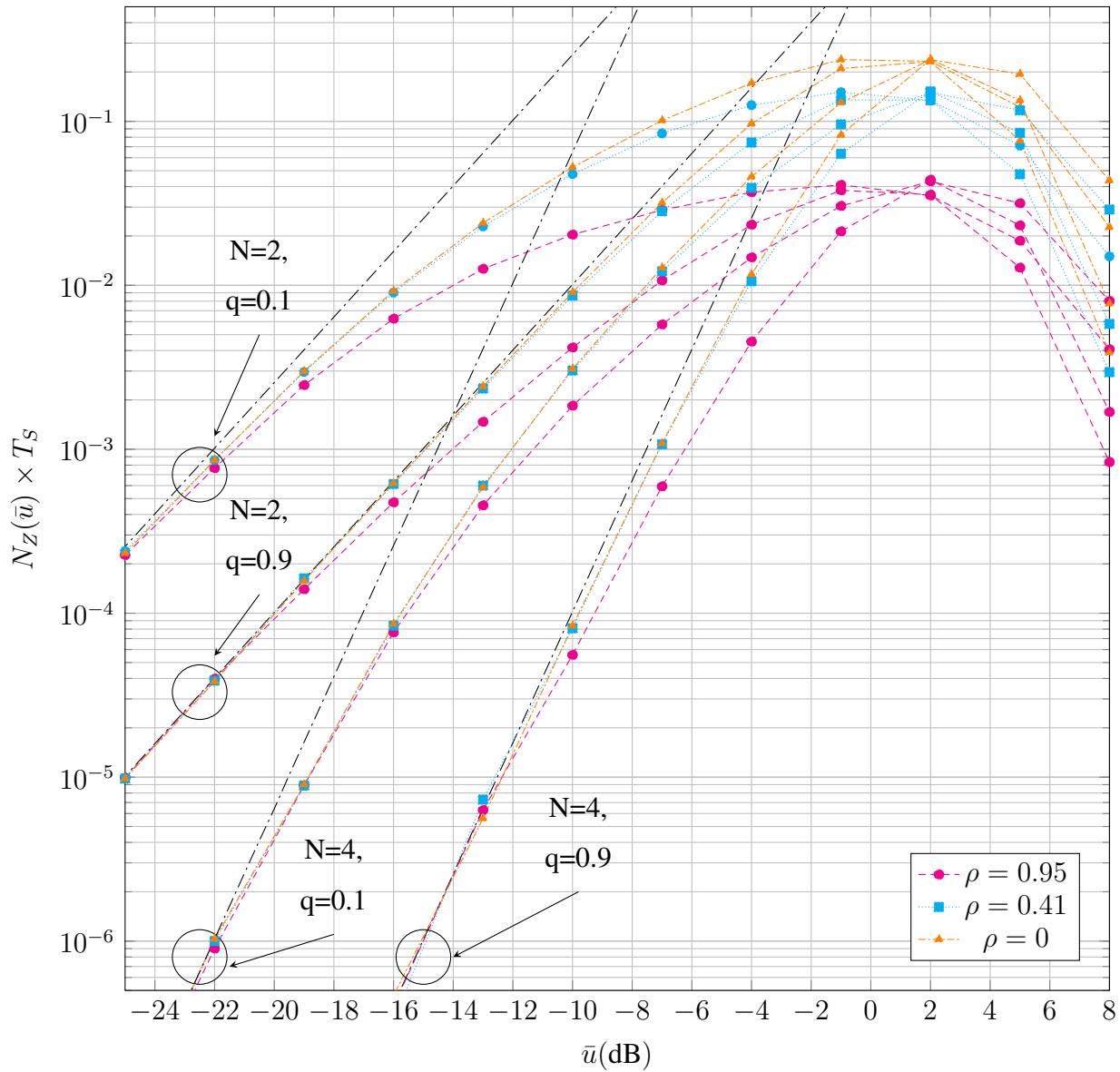

In Fig. \ref{fig_csc_rayleigh}, we show the normalized LCR of the output SNR in colocated diversity SC combiners with different numbers of antennas affected by IID Rayleigh fading. It is shown how high correlation between samples of our discrete processes leads to lower LCR, as it could have been forecast. We also observe how exact curves tend to approach to the asymptotic straight line for low $\bar{u}$ and that, the lower is the correlation, the higher values of $\bar{u}$ for which the asymptotic straight line accurately approximates the exact curve. In this scenario the increase of number of colocated antennas $N$ makes the asymptotic behavior be steeper. Finally, the markers on these curves, and on all the following curves, correspond to simulations points which validate our analysis.

Fig. \ref{fig_csc_hoyt} shows simulation results of normalized LCR of the output SNR in 2- and 4-antenna SC receivers, but IID Hoyt fading is assumed this time. We plot simulated curves to show that our closed-form straight line asymptotic analysis is accurate for low $\bar{u}$. We observe that the value of $q$, the shape parameter of the Hoyt distribution, moves the asymptote location, but not its slope. An explanation for this can be found if we take the Hoyt normalized power PDF, then we attend to (\ref{scunivariate}), and apply the asymptotic analysis described in previous section. We obtain that

\begin{equation}
\log\bigl( \bar{N}_Z(\bar{u})\bigr) |_{\bar{u}\rightarrow 0}\sim N\log\Bigl( \frac{1+q^2}{2q} \Bigr)+ N\log(\bar{u}).
\end{equation}

So the slope is given by the number of branches and the location by $q$ and the number of branches as well.

By setting $q=1$, we obtain the asymptotes in Fig. \ref{fig_csc_rayleigh}, since the Rayleigh distribution equals the $q=1$ Hoyt distribution.

\begin{figure}[htb!]
\centering
\pgfplotsset{every axis/.append style={
xmin={-20},
xmax={10},
ymax={1},
ymin={0.000007},
%ymin={0.00000001},
%extra description/.code={
%\node at (0.58,0.47) {\small{$\Delta T_{DF}$}};
%\node at (0.66,0.38) {\small{$\Delta T$}};
%}
}}
\begin{tikzpicture}[scale=1]
\begin{semilogyaxis}[
width=16cm,
height=16cm,
xlabel=$\bar{u}(\text{dB})$,
ylabel=$N_Z(\bar{u})\times T_S$,
grid=both,
legend entries={$\rho = [0.95\text{ }0.41\text{ }0]$,$\rho=[0\text{ }0.41\text{ }0.95]$},
legend style={legend pos=south east}
]

\addplot[color=magenta, mark=*] table[x=x,y=y] {Legend.dat};
\addplot[color=cyan, mark=square*] table[x=x,y=y] {Legend.dat};

\addplot[color=magenta] table[x=x,y=y] {c1.dat};
\addplot[color=cyan] table[x=x,y=y] {c2.dat};

\addplot[color=magenta, mark=*, only marks] table[x=x,y=y] {c3.dat};
\addplot[color=cyan, mark=square*, only marks] table[x=x,y=y] {c4.dat};

\addplot[color=black, dash pattern=on 6pt off 2pt on 1pt off 2pt] table[x=x,y=y] {c5.dat};

\end{semilogyaxis}
\end{tikzpicture}
\caption{Normalized LCR of the output SNR in a colocated 3-antenna SC combiner in InID Rayleigh scenario. The mean SNRs on each InID Rayleigh are $\Omega$, $\Omega/2$ and $\Omega/4$ respectively}\label{fig_csc_rayleigh_inid}
\end{figure}
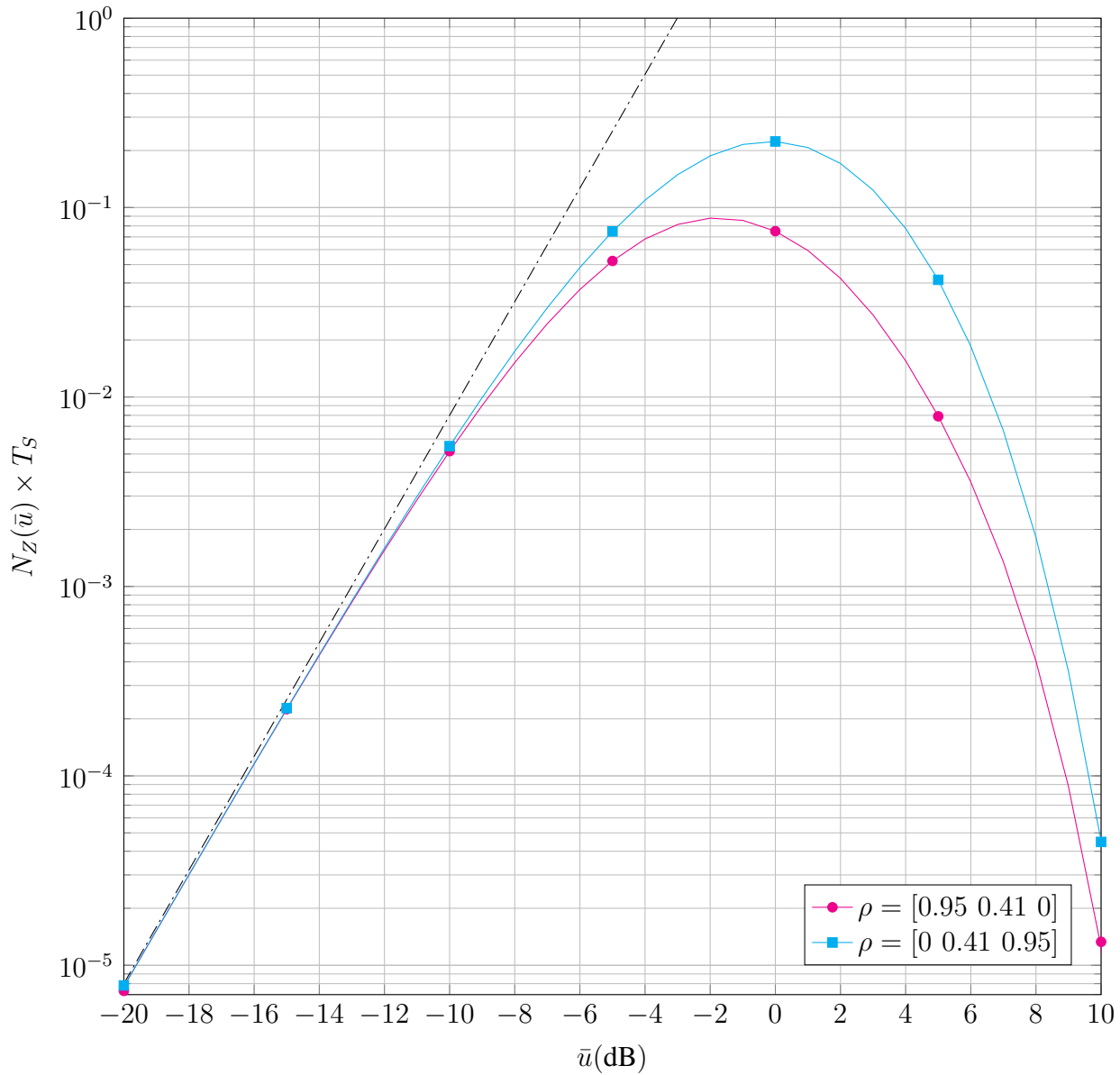

We show a InID Rayleigh colocated SC scenario in Fig. \ref{fig_csc_rayleigh_inid}. We impose to the first branch the mean SNR $\Omega$, which we also use to normalize the level $u$, $\bar{u}=u/\Omega$, and then, fractions of it to the others, concretely $1/2$ and $1/4$. We can observe how the LCR is lower when the most powerful branch, i.e. the most selected, is also highly correlated in time. The expression for this InID Rayleigh SC asymptote reduces to

\begin{equation}
\log\bigl(\bar{N}_Z(\bar{u})\bigr) |_{\bar{u}\rightarrow 0}\sim N\log(\bar{u})-\log\biggl(\prod^{N-1}_{k=0} \beta_k\biggr),
\end{equation}

where $\beta_k$ is the ratio of the mean SNR on each branch and the mean SNR used to normalize $u$, $\beta_k=\bar{Z}_k/\Omega$, i.e. in this case $\beta_0=1$, $\beta_1=1/2$, $\beta_2=1/4$.

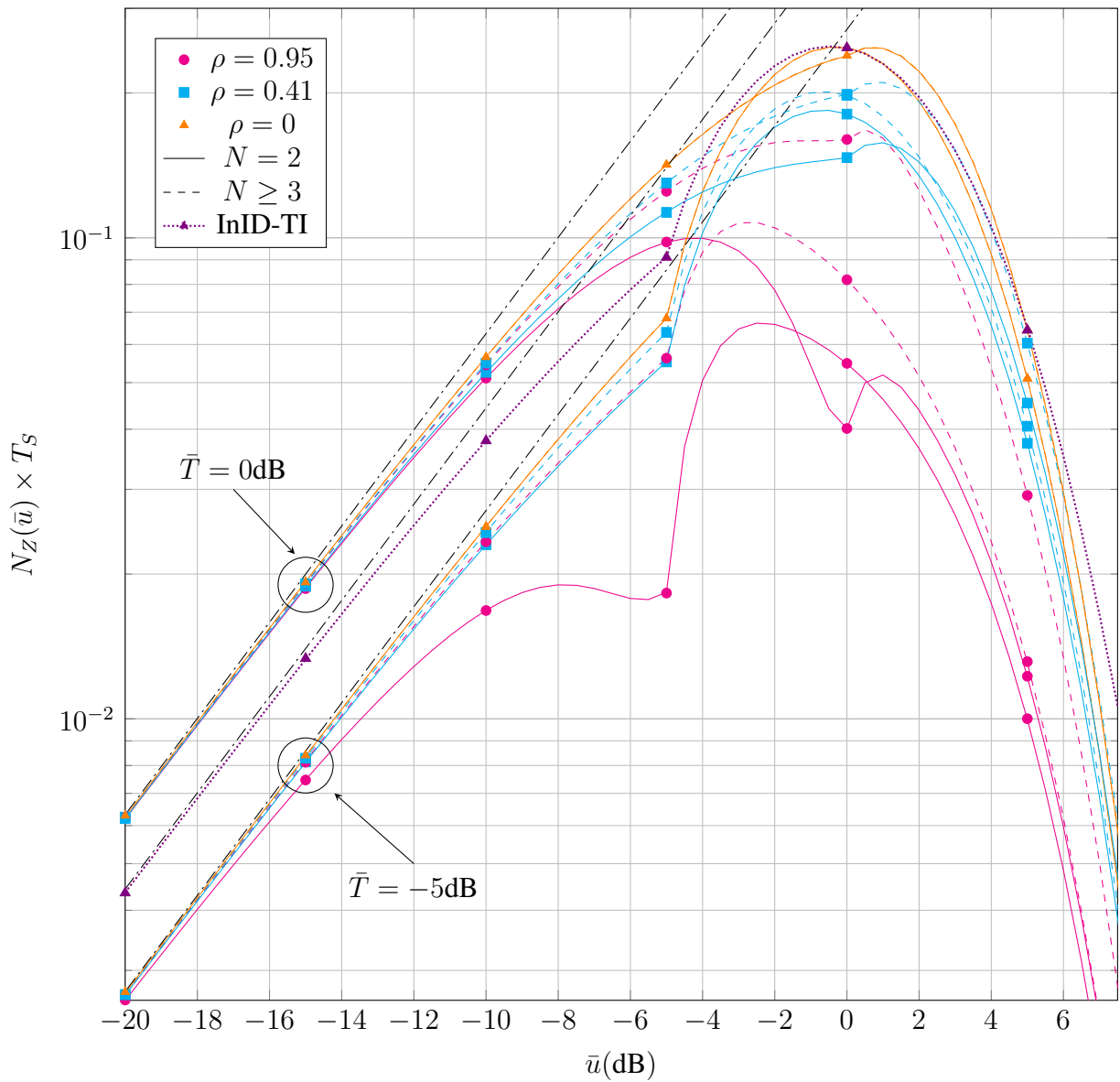
\begin{figure}[htb!]
\centering
\pgfplotsset{every axis/.append style={
xmin={-20},
xmax={7.5},
ymax={0.3},
ymin={0.0026},
extra description/.code={
%\node at (0.58,0.47) {\small{$\Delta T_{DF}$}};
%\node at (0.6,0.65) {\small{$\Delta T$}};
}
}}
\begin{tikzpicture}[scale=1]
\begin{semilogyaxis}[
width=16cm,
height=16cm,
xlabel=$\bar{u}(\text{dB})$,
ylabel=$N_Z(\bar{u})\times T_S$,
grid=both,
legend entries={$\rho = 0.95$,$\rho = 0.41$,$\rho = 0$,$N=2$,$N\geq 3$, InID-TI},
legend style={legend pos=north west}
]

\addplot[color=magenta, only marks, mark=*] table[x=x,y=y] {Legend.dat};
\addplot[color=cyan, only marks, mark=square*] table[x=x,y=y] {Legend.dat};
\addplot[color=orange, only marks, mark=triangle*] table[x=x,y=y] {Legend.dat};

\addplot[color=black] table[x=x,y=y] {Legend.dat};
\addplot[color=black, dashed] table[x=x,y=y] {Legend.dat};

\addplot[color=violet, densely dotted, thick, mark=triangle*] table[x=x,y=y] {Legend.dat};

%%%%%%%%%%%%%%%%%%%%%%%%%%%%%%%%%%%%%%%%%%%%%%%%%%%%%%%%%%%%%%%%%%%%%%%%%%%%%%%%%%%%%%%%%%%%%%%%%%%%%%%%%
\addplot[color=magenta] table[x=x,y=y] {d1.dat};
\addplot[color=magenta, dashed] table[x=x,y=y] {d2.dat};
\addplot[color=magenta] table[x=x,y=y] {d3.dat};
\addplot[color=magenta, dashed] table[x=x,y=y] {d4.dat};

\addplot[color=magenta, only marks, mark=*] table[x=x,y=y] {d5.dat};
\addplot[color=cyan, only marks, mark=square*] table[x=x,y=y] {d6.dat};
\addplot[color=orange, only marks, mark=triangle*] table[x=x,y=y] {d7.dat};

\addplot[color=magenta, only marks, mark=*] table[x=x,y=y] {d8.dat};
\addplot[color=cyan, only marks, mark=square*] table[x=x,y=y] {d9.dat};
\addplot[color=orange, only marks, mark=triangle*] table[x=x,y=y] {d10.dat};

\addplot[color=cyan] table[x=x,y=y] {d11.dat};
\addplot[color=cyan, dashed] table[x=x,y=y] {d12.dat};
\addplot[color=orange] table[x=x,y=y] {d13.dat};
\addplot[color=orange, dashed] table[x=x,y=y] {d14.dat};

\addplot[color=black, dash pattern=on 6pt off 2pt on 1pt off 2pt] table[x=x,y=y] {d15.dat};

%%%%%%%%%%%%%%%%%%%%%%%%%%%%%%%%%%%%%%%%%%%%%%%%%%%%%%%%%%%%%%%%%%%%%%%%%%%%%%%%%%%%%%%%%%%%%%%%%%%%%%%%%

\addplot[color=magenta, only marks, mark=*] table[x=x,y=y] {d16.dat};
\addplot[color=cyan, only marks, mark=square*] table[x=x,y=y] {d17.dat};
\addplot[color=orange, only marks, mark=triangle*] table[x=x,y=y] {d18.dat};

\addplot[color=magenta, only marks, mark=*] table[x=x,y=y] {d19.dat};
\addplot[color=cyan, only marks, mark=square*] table[x=x,y=y] {d20.dat};
\addplot[color=orange, only marks, mark=triangle*] table[x=x,y=y] {d21.dat};

\addplot[color=cyan] table[x=x,y=y] {d22.dat};
\addplot[color=cyan, dashed] table[x=x,y=y] {d23.dat};
\addplot[color=orange] table[x=x,y=y] {d24.dat};
\addplot[color=orange, dashed] table[x=x,y=y] {d25.dat};

\addplot[color=black, dash pattern=on 6pt off 2pt on 1pt off 2pt] table[x=x,y=y] {d26.dat};

%%%%%%%%%%%%%%%%%%%%%%%%%%%%%%%%%%%%%%%%%%%%%%%%%%%%%%%%%%%%%%%%%%%%%%%%%%%%%%%%%%%%%%%%%%%%%%%%%%%%%%%%%

\addplot[color=violet, , densely dotted, thick] table[x=x,y=y] {d27.dat};
\addplot[color=violet, only marks, mark=triangle*, thick] table[x=x,y=y] {d28.dat};

\addplot[color=black, dash pattern=on 6pt off 2pt on 1pt off 2pt] table[x=x,y=y] {d29.dat};

\draw [stealth-] (axis cs:-15.3,0.022) -- (axis cs:-17,0.03) node[above]{$\bar{T}=0$dB};
\draw[] (axis cs:-15,0.019) circle (0.4cm);

\draw [stealth-] (axis cs:-14.2,0.007) -- (axis cs:-12,0.005) node[below]{$\bar{T}=-5$dB};
\draw[] (axis cs:-15,0.008) circle (0.4cm);

\end{semilogyaxis}
\end{tikzpicture}
\caption{Normalized LCR of the output SNR in colocated SSC combiners in IID-AC Rayleigh scenarios and an additional InID-TI scenario.}\label{fig_cssc_rayleigh}
\end{figure}

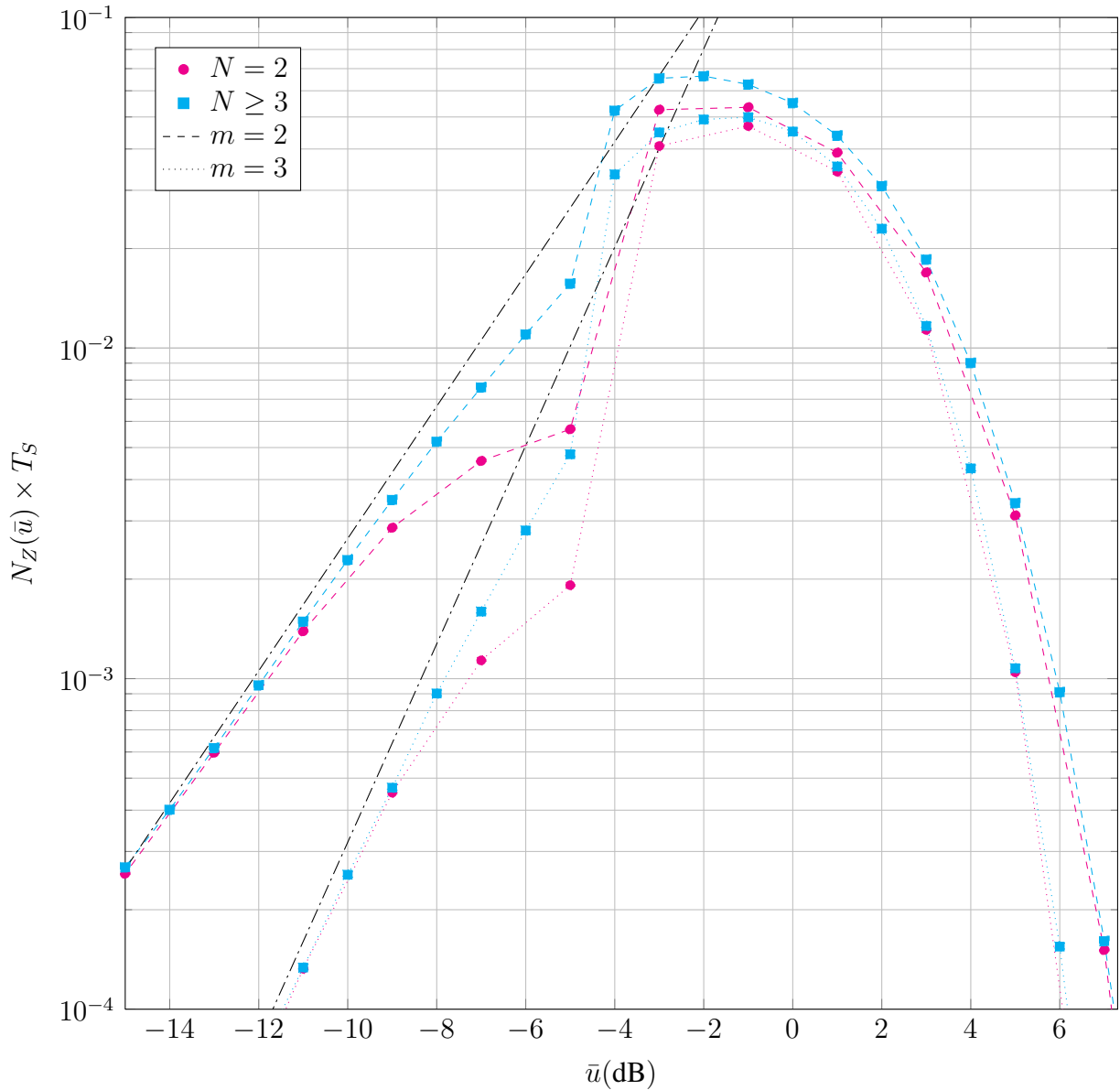
\begin{figure}[htb!]
\centering
\pgfplotsset{every axis/.append style={
xmin={-15},
xmax={7.3},
ymax={0.1},
ymin={0.0001},
extra description/.code={
%\node at (0.58,0.47) {\small{$\Delta T_{DF}$}};
%\node at (0.7,0.55) {\small{$\Delta T$}};
}
}}
\begin{tikzpicture}[scale=1]
\begin{semilogyaxis}[
width=16cm,
height=16cm,
xlabel=$\bar{u}(\text{dB})$,
ylabel=$N_Z(\bar{u})\times T_S$,
grid=both,
legend entries={$N=2$,$N\geq 3$,$m=2$,$m=3$},
legend style={legend pos=north west}
]
\addplot[color=magenta, only marks, mark=*] table[x=x,y=y] {Legend.dat};
\addplot[color=cyan, only marks, thick, mark=square*] table[x=x,y=y] {Legend.dat};

\addplot[dashed] table[x=x,y=y] {Legend.dat};
\addplot[dotted] table[x=x,y=y] {Legend.dat};

%%%%%%%%%%%%%%%%%%%%%%%%%%%%%%%%%%%%%%%%%%%%%%%%%%%%%%%%%%%%%%%%%%%%%%%%%%%%%%%%%%%%%%%%%%%%%%%%%%%%%%%%%

\addplot[color=magenta, dashed, mark=*] table[x=x,y=y] {e1.dat};
\addplot[color=cyan, dashed, mark=square*] table[x=x,y=y] {e2.dat};

\addplot[color=magenta, dotted, mark=*] table[x=x,y=y] {e3.dat};
\addplot[color=cyan, dotted, mark=square*] table[x=x,y=y] {e4.dat};

\addplot[color=black, dash pattern=on 6pt off 2pt on 1pt off 2pt] table[x=x,y=y] {e5.dat};
\addplot[color=black, dash pattern=on 6pt off 2pt on 1pt off 2pt] table[x=x,y=y] {e6.dat};

\end{semilogyaxis}
\end{tikzpicture}
\caption{Comparison of simulation results and asymptotic behaviour of normalized LCR of the output SNR in colocated SSC combiners with $2$ and $3$ or more antennas and $\bar{T}=-5$ dB in $m=2$ and $m=3$ IID Nakagami-m with $\rho=0.95$ scenarios.}\label{fig_cssc_nakagamim}
\end{figure}

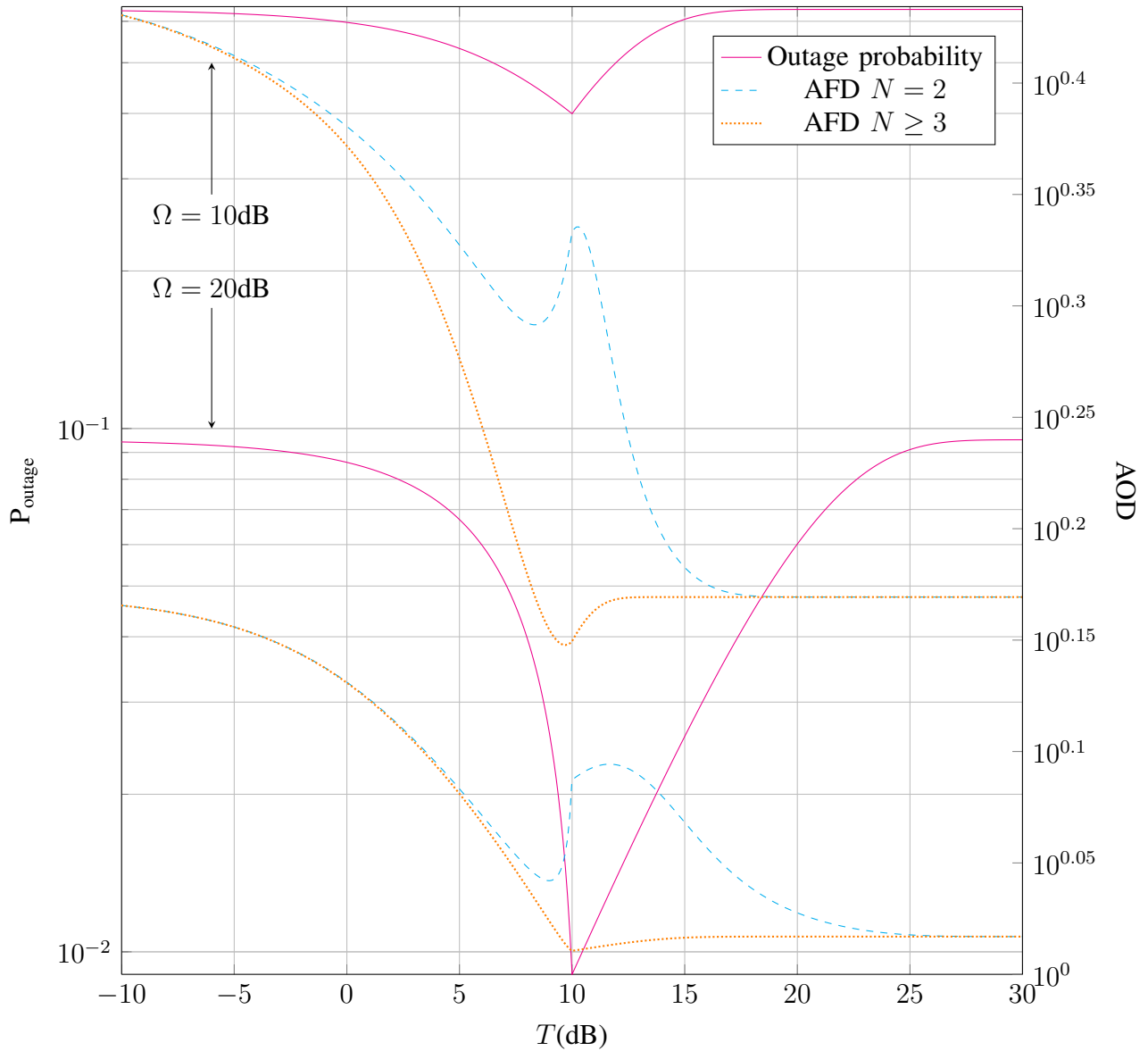
\begin{figure}[htb!]
\centering
\begin{tikzpicture}
  \begin{semilogyaxis}[
  	width=15cm,
	height=16cm,
    xmin = -10, xmax = 30,
    ymin = 0.009056,
    ymax = 0.64,
    %ymin=1e-4,
    %ymax=1e-2,
    axis y line* = left, % the '*' avoids arrow heads
    xlabel = $T$(dB),
    ylabel = $\text{P}_{\text{outage}}$,
    grid=both,
	legend entries={Outage probability, AFD $N=2$, AFD $N\geq 3$},
	legend style={legend pos=north east}]
    \addplot[color=magenta] table[x=x,y=y]{f1.dat};
    \addplot[color=cyan, dashed] table[x=x,y=y]{Legend.dat};
    \addplot[color=orange, thick, densely dotted] table[x=x,y=y]{Legend.dat};
    \addplot[color=magenta] table[x=x,y=y]{f2.dat};
    \draw [stealth-] (axis cs:-6,5e-1) -- (axis cs:-6,2.8e-1) node[below]{$\Omega=10$dB};
    \draw [stealth-] (axis cs:-6,1e-1) -- (axis cs:-6,1.7e-1) node[above]{$\Omega=20$dB};
  \end{semilogyaxis}
  \begin{semilogyaxis}[
  	width=15cm,
	height=16cm,
    xmin = -10, xmax = 30,
    ymin = 1,
    ymax = 13,
    hide x axis,
    hide y axis]
    \addplot[color=cyan, dashed] table[x=x,y=y]{f3.dat};
    \addplot[color=orange, thick, densely dotted] table[x=x,y=y]{f4.dat};
    \addplot[color=cyan, dashed] table[x=x,y=y]{f5.dat};
    \addplot[color=orange, thick, densely dotted] table[x=x,y=y]{f6.dat};
  \end{semilogyaxis}
  \pgfplotsset{every axis y label/.append style={rotate=180,yshift=0.0cm}}
  \begin{semilogyaxis}[
	width=15cm,
	height=16cm,
	ymin=1,
	ymax = 13,
    hide x axis,
	axis y line*=right,
    ylabel={AOD}
    ]
  \end{semilogyaxis}
\end{tikzpicture}
\caption{Outage probability and Average Outage Duration (AOD) (outage level considered is $10$ dB) against the chosen switching threshold $T$ in colocated SSC combiners with $N=2$ or $N\geq 3$ IID Rayleigh diversity sources with per-branch mean SNR $=10$, $20$ dB and $\rho = 0.95$.}\label{fig_cssc_thresholdvsoutage}
\end{figure}

We plot on Fig. \ref{fig_cssc_rayleigh} the normalized LCR of the output SNR of colocated SSC receivers with $N=2$ and $N\geq 3$ antennas in IID Rayleigh fading conditions with switching thresholds $\bar{T}=-5$ dB and $0$ dB. There are exact curves for different correlation cases, whose validity is one more time verified by simulation points. Their straight line asymptotic behavior for low $\bar{u}$ is exhibited. The different number of antennas in this kind of combiners does not change their asymptote, while the switching threshold $T$ changes its location but not its slope. The $N$-dependence is more noticeable in the central parts of the graph near $\bar{T}$. This is because, as we have observed in the asymptotic analysis, only the univariate CDF takes part for low $\bar{u}$ and this one is the same no matter $N$, see (\ref{is-ssc_univariate}). And for high $\bar{u}$, attending to (\ref{is-ssc-bivariate}) common new terms add up to the N-dependent term defined in (\ref{psibivarCDF}). Moreover, we made a difference on the bivariate CDF regarding $N$ because the correlation effect this CDF experiences when returning to the original diversity branch after a second switch or switching to a third branch, see (\ref{psibivarCDF}). That is why we cannot see the $N=2$ and $\rho = 0$ curve, the $N\geq 3$ and $\rho = 0$ curve is the same and has been plotted over the $N=2$ one since there is no difference when different $N$ is available at the receiver and there is no correlation over time on each fading branch. In addition we also plot on Fig. \ref{fig_cssc_rayleigh} a SSC InID-TI example, concretely the dual branch SSC where on one branch we have Rayleigh fading and Hoyt ($q=0.3$) on the other, and we verify its validity with simulation points as well.  

In Fig. \ref{fig_cssc_nakagamim} the results of colocated SSC receivers in IID Nakagami-m environment are plotted. We see how, this time, the $m$ parameter affects on the asymptote slope and the switching threshold to the location. Nevertheless, the number of diversity sources does not change the asymptotic behavior for this kind of combination. An explanation for this can be found if we take the Nakagami-m normalized power PDF, then we attend to (\ref{is-ssc_univariate}), and apply the asymptotic analysis described in previous section. We obtain that

\begin{equation}
\begin{split}
\log\bigl(\bar{N}_Z(\bar{u})\bigr) &|_{\bar{u}\rightarrow 0}\sim\log\Bigl(\frac{m^m}{\Gamma(m)}\frac{(m-1)!}{m!} F_{Z}(\bar{T}|m,\Omega)\Bigr) + m \log(\bar{u}).
\end{split}
\end{equation}

So the slope is given by the parameter $m$ and the location by the switching threshold and $m$ as well.

By setting $m=1$, we obtain the asymptotes in Fig. \ref{fig_cssc_rayleigh}, since the Rayleigh distribution equals the $m=1$ Nakagami-m distribution.

Fig. \ref{fig_cssc_thresholdvsoutage} goes deeper on the performance analysis of SSC systems by adding the study of second order statistics of the outage. We see how the choice of a proper switching threshold in SSC systems is important to reduce the probability of an outage event to occur (left scale on the graph), and that when the threshold equals the outage level the outage probability minimizes. In addition, we observe the Average Outage Duration (AOD) (right scale on the graph) and realize that its minimum is not necessarily reached for the same threshold which minimizes the outage probability. This fact is important to consider for applications where latency of the communication is critical, such as real time communications, where the outage duration bounds the latency limits. Finally, the most important behavior we observe from this curves is that the average outage duration is higher for the dual branch SSC combiner than for the SSC combiner with a third branch or more, which refutes the popular believe that SSC systems do not benefit from more than two diversity branches \cite{sscsystemmodel}.

\begin{figure}[htb!]
\centering
\pgfplotsset{every axis/.append style={
xmin={-25},
xmax={8.7},
ymax={0.3},
ymin={0.001},
extra description/.code={
%\node at (0.58,0.47) {\small{$\Delta T_{DF}$}};
%\node at (0.6,0.65) {\small{$\Delta T$}};
}
}}
\begin{tikzpicture}[scale=1]
\begin{semilogyaxis}[
width=16cm,
height=16cm,
xlabel=$\bar{u}(\text{dB})$,
ylabel=$N_Z(\bar{u})\times T_S$,
grid=both,
legend entries={$\rho = 0.95$,$\rho = 0.41$,$\rho = 0$,$\bar{T}_{DF}=-5\text{dB}$,$\bar{T}_{DF}=+5\text{dB}$},
legend style={legend pos=south west}
]

\addplot[color=magenta, only marks, mark=*] table[x=x,y=y] {Legend.dat};
\addplot[color=cyan, only marks, mark=square*] table[x=x,y=y] {Legend.dat};
\addplot[color=orange, only marks, mark=triangle*] table[x=x,y=y] {Legend.dat};

\addplot[color=black] table[x=x,y=y] {Legend.dat};

\addplot[color=black, dashed] table[x=x,y=y] {Legend.dat};

%%%%%%%%%%%%%%%%%%%%%%%%%%%%%%%%%%%%%%%%%%%%%%%%%%%%%%%%%%%%%%%%%%%%%%%%%%%%%%%%%%%%%%%%%%%%%%%%%%%%%%%%%

\addplot[color=magenta] table[x=x,y=y] {g1.dat};

\addplot[color=magenta, dashed] table[x=x,y=y] {g2.dat};

\addplot[color=cyan] table[x=x,y=y] {g3.dat};

\addplot[color=cyan, dashed] table[x=x,y=y] {g4.dat};

\addplot[color=orange] table[x=x,y=y] {g5.dat};

\addplot[color=orange, dashed] table[x=x,y=y] {g6.dat};

%%%%%%%%%%%%%%%%%%%%%%%%%%%%%%%%%%%%%%%%%%%%%%%%%%%%%%%%%%%%%%%%%%%%%%%%%%%%%%%%%%%%%%%%%%%%%%%%%%%%%%%%%

\addplot[color=magenta, only marks, mark=*] table[x=x,y=y] {g7.dat};

\addplot[color=magenta, only marks, mark=*] table[x=x,y=y] {g8.dat};

\addplot[color=cyan, only marks, mark=square*] table[x=x,y=y] {g9.dat};

\addplot[color=cyan, only marks, mark=square*] table[x=x,y=y] {g10.dat};

\addplot[color=orange, only marks, mark=triangle*] table[x=x,y=y] {g11.dat};

\addplot[color=orange, only marks, mark=triangle*] table[x=x,y=y] {g12.dat};

%%%%%%%%%%%%%%%%%%%%%%%%%%%%%%%%%%%%%%%%%%%%%%%%%%%%%%%%%%%%%%%%%%%%%%%%%%%%%%%%%%%%%%%%%

\addplot[color=black, dash pattern=on 6pt off 2pt on 1pt off 2pt] table[x=x,y=y] {g13.dat};

\addplot[color=black, dash pattern=on 6pt off 2pt on 1pt off 2pt] table[x=x,y=y] {g14.dat};

\addplot[color=black, dash pattern=on 6pt off 2pt on 1pt off 2pt] table[x=x,y=y] {g15.dat};

\addplot[color=black, dash pattern=on 6pt off 2pt on 1pt off 2pt] table[x=x,y=y] {g16.dat};

\addplot[color=black, dash pattern=on 6pt off 2pt on 1pt off 2pt] table[x=x,y=y] {g17.dat};

\addplot[color=black, dash pattern=on 6pt off 2pt on 1pt off 2pt] table[x=x,y=y] {g18.dat};

\end{semilogyaxis}
\end{tikzpicture}
\caption{Normalized LCR of the output SNR in IID Rayleigh dual hop dual branch ($N=2$) Opportunistic Relaying scenario.}\label{fig_dsc_rayleigh}
\end{figure}
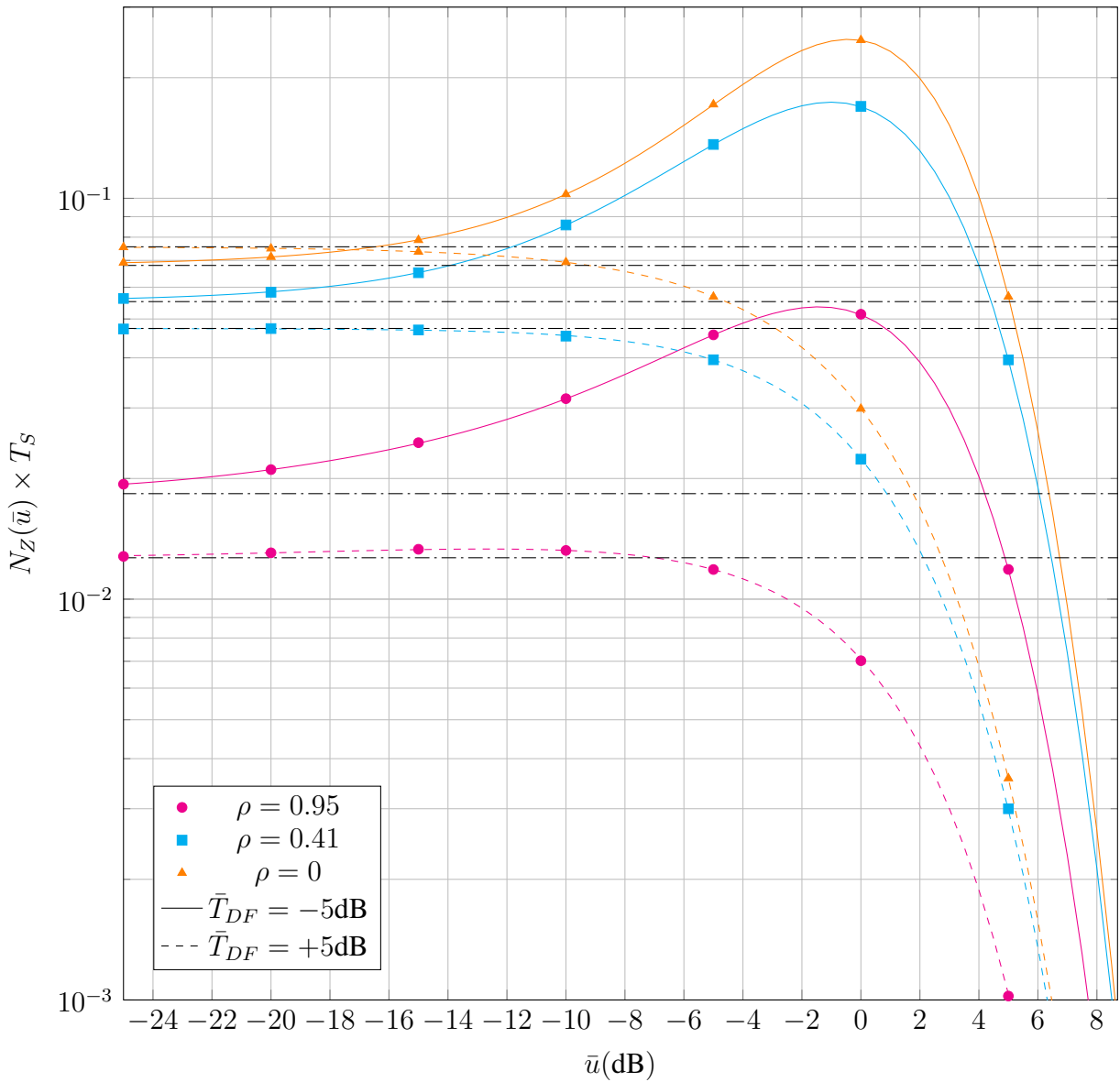

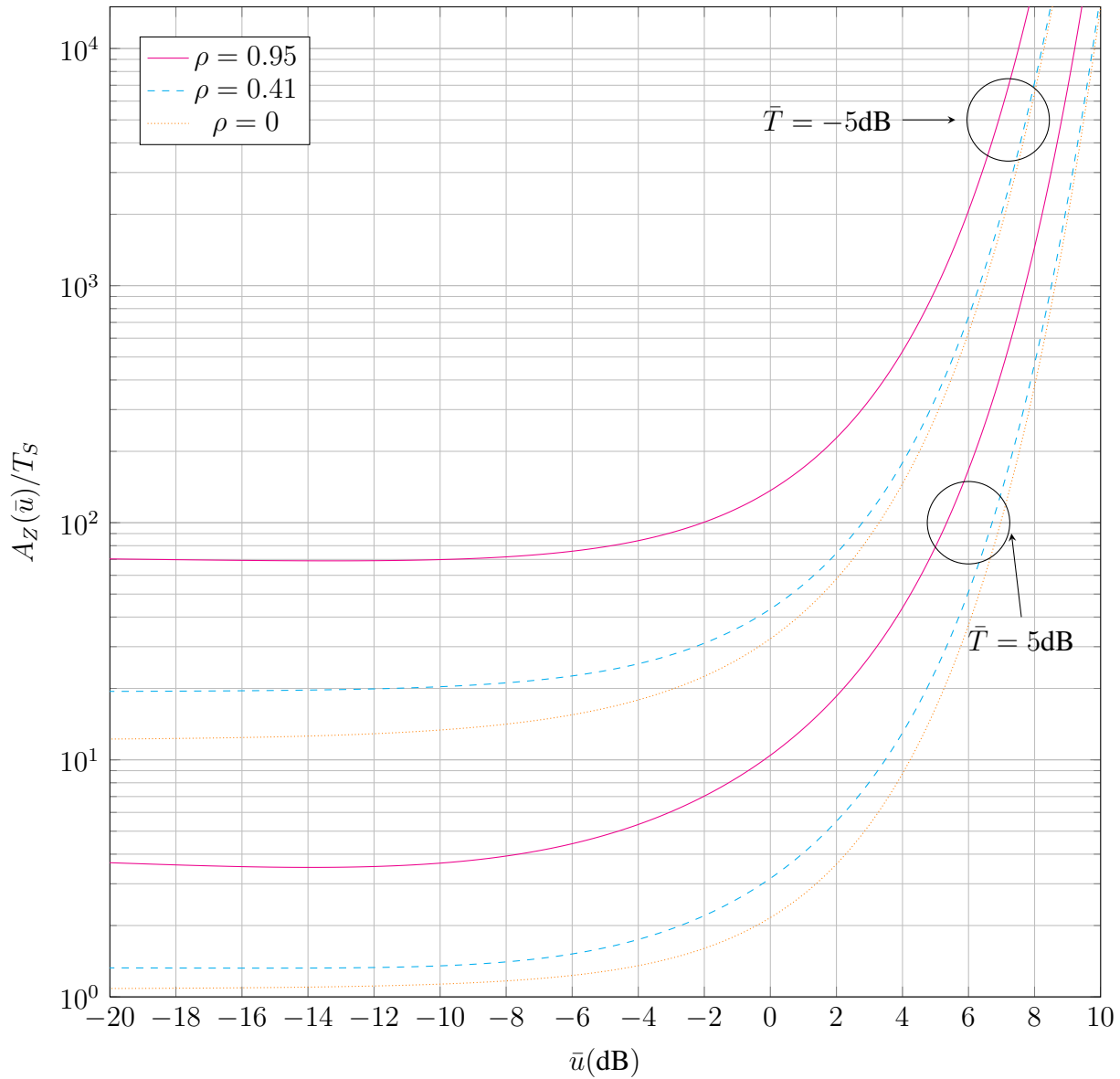
\begin{figure}[htb!]
\centering
\pgfplotsset{every axis/.append style={
xmin={-20},
xmax={10},
ymax={15000},
ymin={1},
extra description/.code={
%\node at (0.58,0.47) {\small{$\Delta T_{DF}$}};
%\node at (0.6,0.65) {\small{$\Delta T$}};
}
}}
\begin{tikzpicture}[scale=1]
\begin{semilogyaxis}[
width=16cm,
height=16cm,
xlabel=$\bar{u}(\text{dB})$,
ylabel=$A_Z(\bar{u})/T_S$,
grid=both,
legend entries={$\rho = 0.95$,$\rho = 0.41$,$\rho = 0$},
legend style={legend pos=north west}
]

\addplot[color=magenta] table[x=x,y=y] {Legend.dat};
\addplot[color=cyan, dashed] table[x=x,y=y] {Legend.dat};
\addplot[color=orange, densely dotted] table[x=x,y=y] {Legend.dat};

%%%%%%%%%%%%%%%%%%%%%%%%%%%%%%%%%%%%%%%%%%%%%%%%%%%%%%%%%%%%%%%%%%%%%%%%%%%%%%%%%%%%%%%%%%%%%%%%%%%%%%%%%

\addplot[color=magenta] table[x=x,y=y] {h1.dat};

\addplot[color=magenta] table[x=x,y=y] {h2.dat};

\addplot[color=cyan, dashed] table[x=x,y=y] {h3.dat};

\addplot[color=cyan, dashed] table[x=x,y=y] {h4.dat};

\addplot[color=orange, densely dotted] table[x=x,y=y] {h5.dat};

\addplot[color=orange, densely dotted] table[x=x,y=y] {h6.dat};

\draw [stealth-] (axis cs:5.6,5000) -- (axis cs:4,5000) node[left]{$\bar{T}=-5$dB};
\draw[] (axis cs:7.2,5000) circle (0.6cm);

\draw [stealth-] (axis cs:7.3,90) -- (axis cs:7.6,40) node[below]{$\bar{T}=5$dB};
\draw[] (axis cs:6,100) circle (0.6cm);

\end{semilogyaxis}
\end{tikzpicture}
\caption{Normalized AFD of the output SNR in IID Rayleigh dual hop dual branch ($N=2$) Opportunistic Relaying scenario.}\label{fig_dsc_afd_rayleigh}
\end{figure}

Fig. \ref{fig_dsc_rayleigh} shows the normalized LCR of the SNR at a single antenna receiver from a dual hop dual branch threshold based DF OR scheme when all the links involved in the configuration undergoes IID Rayleigh fading. We plot exact curves verified by simulation points. For this configuration the asymptotic behavior is a horizontal straight line due to threshold based DF mode of operation, which condenses a finite amount of probability at $Z=0$. Attending to (\ref{scunivariate}), (\ref{scbivariate}), (\ref{distributed_univariate}) and (\ref{distributed_bivariate}), in the IID case, we can write the expression for the horizontal asymptote,

\begin{equation}
\begin{split}
\bar{N}_Z(\bar{u})|_{\bar{u}\rightarrow 0}\sim
\bigl( F_{\bar{Z}}(\bar{T}_{DF})\bigr) ^N - \bigl(F_{\bar{\mathbf{Z}}}(\bar{T}_{DF},\bar{T}_{DF}|\rho)\bigr) ^N\text{,}
\end{split}
\end{equation}

where $F_{\bar{Z}}(\bar{z})$ and $F_{\bar{\mathbf{Z}}}(\bar{z},\bar{z}|\rho)$ represent the univariate and bivariate CDF of the IID fading on each branch.

We observe that in this particular scenario the correlation affects the asymptotic behavior, in particular on its location: low correlation rises the asymptote. Another parameter which changes the asymptote location is the decoding threshold $T_{DF}$.

In Fig. \ref{fig_dsc_afd_rayleigh} we plot the normalized AFD for the same scenarios as Fig. \ref{fig_dsc_rayleigh}. The AFD does not tend to $1$ for low $\bar{u}$ as we had foreseen in previous section due to the condensed probability at $Z=0$ caused by the threshold based DF relays. Higher $\bar{T}_{DF}$ means higher probability of the relays not forwarding the signal, so the AFD rises. The higher temporal correlation also makes the AFD rise. These two phenomenona are evidenced in the figure.

\begin{figure}[htb!]
\centering
\pgfplotsset{every axis/.append style={
xmin={-25},
xmax={5},
ymax={0.5},
ymin={0.0085},
extra description/.code={
%\node at (0.58,0.47) {\small{$\Delta T_{DF}$}};
%\node at (0.6,0.65) {\small{$\Delta T$}};
}
}}
\begin{tikzpicture}[scale=1]
\begin{semilogyaxis}[
width=16cm,
height=16cm,
xlabel=$\bar{u}(\text{dB})$,
ylabel=$N_Z(\bar{u})\times T_S$,
grid=both,
legend entries={$\rho = 0.95$,$\rho = 0.41$,$\rho = 0$,$\bar{T}=-5\text{dB, }\bar{T}_{DF}=-8\text{dB}$,$\bar{T}=0\text{dB, }\bar{T}_{DF}=-3\text{dB}$,$\bar{T}=5\text{dB, }\bar{T}_{DF}=2\text{dB}$},
legend style={legend pos=south west}
]

\addplot[color=magenta, only marks, mark=*] table[x=x,y=y] {Legend.dat};
\addplot[color=cyan, only marks, mark=square*] table[x=x,y=y] {Legend.dat};
\addplot[color=orange, only marks, mark=triangle*] table[x=x,y=y] {Legend.dat};

\addplot[color=black] table[x=x,y=y] {Legend.dat};
\addplot[dashed] table[x=x,y=y] {Legend.dat};
\addplot[densely dotted] table[x=x,y=y] {Legend.dat};

%%%%%%%%%%%%%%%%%%%%%%%%%%%%%%%%%%%%%%%%%%%%%%%%%%%%%%%%%%%%%%%%%%%%%%%%%%%%%%%%%%%%%%%%%%%%%%%%%%%%%%%%%

\addplot[color=magenta, dashed] table[x=x,y=y] {i1.dat};
\addplot[color=magenta] table[x=x,y=y] {i2.dat};
\addplot[color=magenta, densely dotted] table[x=x,y=y] {i3.dat};

\addplot[color=cyan, dashed] table[x=x,y=y] {i4.dat};
\addplot[color=cyan] table[x=x,y=y] {i5.dat};
\addplot[color=cyan, densely dotted] table[x=x,y=y] {i6.dat};

\addplot[color=orange, dashed] table[x=x,y=y] {i7.dat};
\addplot[color=orange] table[x=x,y=y] {i8.dat};
\addplot[color=orange, densely dotted] table[x=x,y=y] {i9.dat};

%%%%%%%%%%%%%%%%%%%%%%%%%%%%%%%%%%%%%%%%%%%%%%%%%%%%%%%%%%%%%%%%%%%%%%%%%%%%%%%%%%%%%%%%%%%%%%%%%%%%%%%%%

\addplot[color=magenta, only marks, mark=*] table[x=x,y=y] {i10.dat};
\addplot[color=magenta, only marks, mark=*] table[x=x,y=y] {i11.dat};
\addplot[color=magenta, only marks, mark=*] table[x=x,y=y] {i12.dat};

\addplot[color=cyan, only marks, mark=square*] table[x=x,y=y] {i13.dat};
\addplot[color=cyan, only marks, mark=square*] table[x=x,y=y] {i14.dat};
\addplot[color=cyan, only marks, mark=square*] table[x=x,y=y] {i15.dat};

\addplot[color=orange, only marks, mark=triangle*] table[x=x,y=y] {i16.dat};
\addplot[color=orange, only marks, mark=triangle*] table[x=x,y=y] {i17.dat};
\addplot[color=orange, only marks, mark=triangle*] table[x=x,y=y] {i18.dat};

\addplot[color=black, dash pattern=on 6pt off 2pt on 1pt off 2pt] table[x=x,y=y] {i19.dat};
\addplot[color=black, dash pattern=on 6pt off 2pt on 1pt off 2pt] table[x=x,y=y] {i20.dat};
\addplot[color=black, dash pattern=on 6pt off 2pt on 1pt off 2pt] table[x=x,y=y] {i21.dat};

\end{semilogyaxis}
\end{tikzpicture}
\caption{Normalized LCR of the output SNR in distributed SSC combiners in IID Rayleigh scenario ($\rho=0.05$).}\label{fig_dssc_rayleigh}
\end{figure}
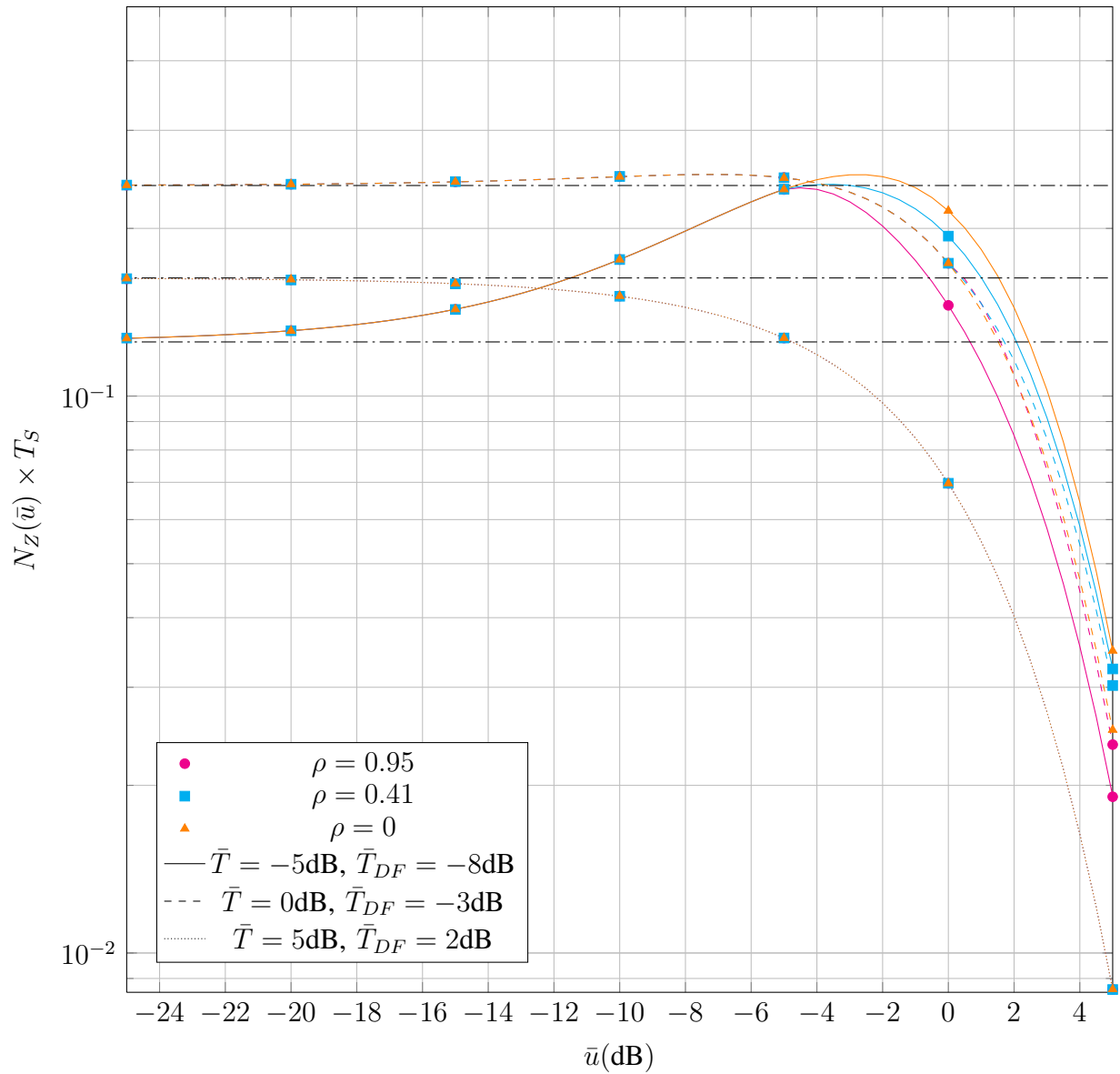

In Fig. \ref{fig_dssc_rayleigh}, it is shown the normalized LCR of the SNR at a single antenna receiver from a distributed SSC dual hop threshold based DF  relaying scheme when all the links involved in the configuration undergoes IID Rayleigh fading for any $N$. We plot exact curves verified by simulation points. For this configuration the asymptotic behavior is again a horizontal straight line due to threshold based DF mode of operation. Concretely, attending to (\ref{ssc_distributed_univariate}), (\ref{ssc_distributed_bivariate}) and (\ref{distributed_univariate}), in the IID case,

\begin{equation}
\bar{N}_Z(\bar{u})|_{\bar{u}\rightarrow 0}\sim F_{\bar{Z}}(\bar{T}_{DF})\bigl( 1-F_{\bar{Z}}(\bar{T}_{DF}) \bigr).
\end{equation}

We observe here that the location of the horizontal asymptotes only depends on the decoding threshold $T_{DF}$, but not on the per-branch correlation. In fact, this scheme exhibits little difference for different correlations.

\section{Conclusion}\label{conclusiones}
We derived novel expressions for the LCR and AFD of communication systems based on the concept of switched diversity diversity to achieve full-order spatial diversity. Threshold based techniques lacked from this analytical characterization for these statistics because of the inherent discontinuity of the random processes of interest; we circumvented the limitations of Rice's approach by using an alternative formulation for sampled random process. We achieved to characterize SC and SSC techniques, which are now again subject of interest since the introduction of cooperative diversity, in classic colocated diversity and already mentioned cooperative distributed diversity  scenarios in a discrete time fashion, which matches the actual inherent implementation of these techniques. Furthermore, the SSC schemes lacked of LCR and AFD analytical characterization until now to the best of our knowledge. We also provided another simple manner to elude the discontinuity when analyzing the higher order statistics of threshold-based DF relaying systems.

Our analysis holds for an arbitrary fading distribution either when assuming colocated or distributed diversity branches. In the process of computing the LCR and the AFD, we obtained analytical expressions for the univariate and bivariate CDFs of the output SNR. We also studied the effect of temporal correlation in these scenarios. Moreover, we introduced a common analysis for the asymptotic LCR and AFD in high mean SNR environments which allows to characterize the behavior of the higher order statistics only requiring the availability of the univariate PDF which describes the fading random process. We also discussed the implications of using a number of branches in SSC systems larger than the conventional recommendation of two for wireless systems finding interesting results about how more than two branches improves the average outage duration. Finding that a third branch in SSC systems reduces the average outage duration of these.

% if have a single appendix:
%\appendix[Proof of the Zonklar Equations]
% or
%\appendix  % for no appendix heading
% do not use \section anymore after \appendix, only \section*
% is possibly needed

% use appendices with more than one appendix
% then use \section to start each appendix
% you must declare a \section before using any
% \subsection or using \label (\appendices by itself
% starts a section numbered zero.)
%

%\appendices
%\section{Proof of the First Zonklar Equation}
%Appendix one text goes here.

% you can choose not to have a title for an appendix
% if you want by leaving the argument blank
%\section{}
%Appendix two text goes here.

% use section* for acknowledgment
%\section*{Acknowledgment}
%
%This work is partially supported by the Spanish Government and FEDER under project TEC2011-25473 and the Junta de Andaluc\'{i}a under projects P11-TIC-7109 and P11-TIC-8238.

% Can use something like this to put references on a page
% by themselves when using endfloat and the captionsoff option.
\ifCLASSOPTIONcaptionsoff
  \newpage
\fi

% trigger a \newpage just before the given reference
% number - used to balance the columns on the last page
% adjust value as needed - may need to be readjusted if
% the document is modified later
%\IEEEtriggeratref{8}
% The "triggered" command can be changed if desired:
%\IEEEtriggercmd{\enlargethispage{-5in}}

% references section

% can use a bibliography generated by BibTeX as a .bbl file
% BibTeX documentation can be easily obtained at:
% http://www.ctan.org/tex-archive/biblio/bibtex/contrib/doc/
% The IEEEtran BibTeX style support page is at:
% http://www.michaelshell.org/tex/ieeetran/bibtex/
\bibliographystyle{IEEEtran}
% argument is your BibTeX string definitions and bibliography database(s)
\bibliography{bibfile}

% Generated by IEEEtran.bst, version: 1.13 (2008/09/30)
\begin{thebibliography}{10}
\providecommand{\url}[1]{#1}
\csname url@samestyle\endcsname
\providecommand{\newblock}{\relax}
\providecommand{\bibinfo}[2]{#2}
\providecommand{\BIBentrySTDinterwordspacing}{\spaceskip=0pt\relax}
\providecommand{\BIBentryALTinterwordstretchfactor}{4}
\providecommand{\BIBentryALTinterwordspacing}{\spaceskip=\fontdimen2\font plus
\BIBentryALTinterwordstretchfactor\fontdimen3\font minus
  \fontdimen4\font\relax}
\providecommand{\BIBforeignlanguage}[2]{{%
\expandafter\ifx\csname l@#1\endcsname\relax
\typeout{** WARNING: IEEEtran.bst: No hyphenation pattern has been}%
\typeout{** loaded for the language `#1'. Using the pattern for}%
\typeout{** the default language instead.}%
\else
\language=\csname l@#1\endcsname
\fi
#2}}
\providecommand{\BIBdecl}{\relax}
\BIBdecl

\bibitem{coopdiversity1}
J.~N. Laneman and G.~W. Wornell, ``Distributed space-time-coded protocols for
  exploiting cooperative diversity in wireless networks,'' \emph{IEEE Trans.
  Inf. Theory}, vol.~49, no.~10, pp. 2415--2425, Oct 2003.

\bibitem{coopdiversity2}
J.~N. Laneman, D.~N.~C. Tse, and G.~W. Wornell, ``Cooperative diversity in
  wireless networks: {E}fficient protocols and outage behavior,'' \emph{IEEE
  Trans. Inf. Theory}, vol.~50, no.~12, pp. 3062--3080, Dec 2004.

\bibitem{coopdiversity3}
M.~Uysal, \emph{Cooperative Communications for Improved Wireless Network
  Transmission: Framework for Virtual Antenna Array Applications}.\hskip 1em
  plus 0.5em minus 0.4em\relax Hershey, PA: Information Science Reference,
  2009.

\bibitem{dsscperf5}
M.~K. Jataprolu, D.~S. Michalopoulos, and R.~Schober, ``Colocated and
  distributed switch-and-stay combining: {O}ptimality under switching rate
  constraints,'' \emph{IEEE Trans. Veh. Technol.}, vol.~63, no.~1, pp.
  451--457, Jan 2014.

\bibitem{switcheddiversity1}
P.~A. Anghel and M.~Kaveh, ``Exact symbol error probability of a cooperative
  network in a {R}ayleigh-fading environment,'' \emph{IEEE Trans. Wireless
  Commun.}, vol.~3, no.~5, pp. 1416--1421, Sept 2004.

\bibitem{switcheddiversity2}
A.~Ribeiro, X.~Cai, and G.~B. Giannakis, ``Symbol error probabilities for
  general cooperative links,'' \emph{IEEE Trans. Wireless Commun.}, vol.~4,
  no.~3, pp. 1264--1273, May 2005.

\bibitem{switcheddiversity3}
Y.~Zhao, R.~Adve, and T.~J. Lim, ``Symbol error rate of selection
  amplify-and-forward relay systems,'' \emph{IEEE Commun. Lett.}, vol.~10,
  no.~11, pp. 757--759, November 2006.

\bibitem{switcheddiversity4}
D.~S. Michalopoulos and G.~K. Karagiannidis, ``Performance analysis of single
  relay selection in {R}ayleigh fading,'' \emph{IEEE Trans. Wireless Commun.},
  vol.~7, no.~10, pp. 3718--3724, October 2008.

\bibitem{alouini}
M.~K. Simon and M.-S. Alouini, \emph{Digital communication over fading
  channels}.\hskip 1em plus 0.5em minus 0.4em\relax John Wiley \& Sons, 2005,
  vol.~95.

\bibitem{csifb1}
A.~Bletsas, A.~Khisti, D.~P. Reed, and A.~Lippman, ``A simple cooperative
  diversity method based on network path selection,'' \emph{IEEE J. Sel. Areas
  Commun.}, vol.~24, no.~3, pp. 659--672, March 2006.

\bibitem{dsscintro1}
D.~S. Michalopoulos and G.~K. Karagiannidis, ``Distributed switch and stay
  combining (dssc) with a single decode and forward relay,'' \emph{IEEE
  Communications Letters}, vol.~11, no.~5, pp. 408--410, May 2007.

\bibitem{dsscintro2}
------, ``Two-relay distributed switch and stay combining,'' \emph{IEEE Trans.
  Commun.}, vol.~56, no.~11, pp. 1790--1794, November 2008.

\bibitem{aoraod}
N.~Zlatanov, Z.~Hadzi-Velkov, G.~K. Karagiannidis, and R.~Schober,
  ``Cooperative diversity with mobile nodes: Capacity outage rate and
  duration,'' \emph{IEEE Trans. Inf. Theory}, vol.~57, no.~10, pp. 6555--6568,
  Oct 2011.

\bibitem{dsscperf1}
V.~N.~Q. Bao and H.-Y. Kong, ``Distributed switch and stay combining for
  selection relay networks,'' \emph{IEEE Commun. Lett.}, vol.~13, no.~12, pp.
  914--916, December 2009.

\bibitem{dsscperf3}
M.~Yan, Q.~Chen, X.~Lei, T.~Q. Duong, and P.~Fan, ``Outage probability of
  switch and stay combining in two-way amplify-and-forward relay networks,''
  \emph{IEEE Wireless Commun. Lett.}, vol.~1, no.~4, pp. 296--299, August 2012.

\bibitem{dsscperf4}
L.~Fan, X.~Lei, R.~Q. Hu, and S.~Zhang, ``Distributed two-way switch and stay
  combining with a single amplify-and-forward relay,'' \emph{IEEE Wireless
  Commun. Lett.}, vol.~2, no.~4, pp. 379--382, August 2013.

\bibitem{dsscperf6}
Z.~Paruk and H.~Xu, ``Distributed switch and stay combining with partial relay
  selection and signal space diversity,'' \emph{IET Communications}, vol.~8,
  no.~1, pp. 105--113, Jan 2014.

\bibitem{rice}
S.~O. Rice, \emph{Mathematical analysis of random noise}.\hskip 1em plus 0.5em
  minus 0.4em\relax Murray Hill, NJ: BTL, 1944.

\bibitem{ref_LCRMIMO_1}
C.-D. Iskander and P.~T. Mathiopoulos, ``Analytical level crossing rates and
  average fade durations for diversity techniques in {N}akagami fading
  channels,'' \emph{IEEE Trans. Commun.}, vol.~50, no.~8, pp. 1301--1309, Aug
  2002.

\bibitem{ref_LCRMIMO_2}
N.~C. Beaulieu and X.~Dong, ``{Level crossing rate and average fade duration of
  MRC and EGC diversity in Ricean fading},'' \emph{IEEE Trans. Commun.},
  vol.~51, no.~5, pp. 722--726, May 2003.

\bibitem{ref_LCRMIMO_3}
G.~Fraidenraich, M.~D. Yacoub, and J.~C.~S. Santos~Filho, ``Second-order
  statistics of maximal-ratio and equal-gain combining in {W}eibull fading,''
  \emph{IEEE Commun. Lett.}, vol.~9, no.~6, pp. 499--501, Jun 2005.

\bibitem{letteraoraod}
Z.~Hadzi-Velkov and N.~Zlatanov, ``Outage rates and outage durations of
  opportunistic relaying systems,'' \emph{IEEE Communications Letters},
  vol.~14, no.~2, pp. 148--150, February 2010.

\bibitem{lcrssc1}
L.~Yang and M.-S. Alouini, ``Average level crossing rate and average outage
  duration of switched diversity systems,'' in \emph{Global Telecommunications
  Conference}, vol.~2, Nov 2002.

\bibitem{paperedu}
F.~J. L\'{o}pez-Mart\'{i}nez, E.~Martos-Naya, J.~F. Paris, and
  U.~Fern\'{a}ndez-Plazaola, ``Higher order statistics of sampled fading
  channels with applications,'' \emph{IEEE Trans. Veh. Technol.}, vol.~61,
  no.~7, pp. 3342--3346, Sept 2012.

\bibitem{sscsystemmodel}
H.~Yang and M.-S. Alouini, ``Performance analysis of multi-branch switched
  diversity systems,'' in \emph{IEEE 55th Vehicular Technology Conference},
  vol.~2, 2002.

\bibitem{sscabc}
H.-C. Yang and M.-S. Alouini, ``Markov chains and performance comparison of
  switched diversity systems,'' \emph{IEEE Trans. Commun.}, vol.~52, no.~7, pp.
  1113--1125, July 2004.

\bibitem{tableintegrals}
I.~S. Gradshteyn and I.~M. Ryzhik, \emph{Table of integrals, series and
  products}.\hskip 1em plus 0.5em minus 0.4em\relax Academic Press, 2007.

\bibitem{jakes}
W.~C. Jakes, \emph{Microwave mobile communications}.\hskip 1em plus 0.5em minus
  0.4em\relax Wiley-IEEE Press, 1974.

\end{thebibliography}
\end{document}